\documentclass[twocolumn,
pra
,superscriptaddress
,floatfix
,aps
,10pt
,showpacs]{revtex4-1}

\usepackage{graphicx}
\usepackage{amssymb}
\usepackage{amsmath}
\usepackage{multirow}

\usepackage{array,tabularx}
\usepackage{color}

\newcommand{\vect}[1]{{\mathbf #1}}
\newcommand{\Frac}[2]{\displaystyle\frac{#1}{#2}}
\renewcommand{\k}{{\bf k}}

\begin{document}


\title{Spontaneous patterns in coherently driven polariton
  microcavities}

\author{G. D\'iaz-Camacho}
\affiliation{Departamento de F\'isica Te\'orica de la Materia
  Condensada \& Condensed Matter Physics Center (IFIMAC), Universidad
  Aut\'onoma de Madrid, Madrid 28049, Spain}

\author{C. Tejedor}
\affiliation{Departamento de F\'isica Te\'orica de la Materia
  Condensada \& Condensed Matter Physics Center (IFIMAC), Universidad
  Aut\'onoma de Madrid, Madrid 28049, Spain}

\author{F. M. Marchetti}
\email{francesca.marchetti@uam.es}
\affiliation{Departamento de F\'isica Te\'orica de la Materia
  Condensada \& Condensed Matter Physics Center (IFIMAC), Universidad
  Aut\'onoma de Madrid, Madrid 28049, Spain}

\date{\today}

\begin{abstract}
  We consider a polariton microcavity resonantly driven by two
  external lasers which simultaneously pump both lower and upper
  polariton branches at normal incidence. In this setup, we study the
  occurrence of instabilities of the pump-only solutions towards the
  spontaneous formation of patterns. Their appearance is a consequence
  of the spontaneous symmetry breaking of translational and rotational
  invariance due to interaction induced parametric scattering.
  We observe the evolution between diverse patterns which can be
  classified as single-pump, where parametric scattering occurs at the
  same energy as one of the pumps, and as two-pump, where scattering
  occurs at a different energy. For two-pump instabilities, stripe and
  chequerboard patterns become the dominant steady-state solutions
  because cubic parametric scattering processes are forbidden. This
  contrasts with the single-pump case, where hexagonal patterns are
  the most common arrangements. We study the possibility of
  controlling the evolution between different patterns.
  Our results are obtained within a linear stability analysis and are
  confirmed by finite size full numerical calculations.
\end{abstract}

\pacs{}

\maketitle

\section{Introduction}
In recent years, hybrid matter-light systems such as microcavity
polaritons have been proven ideal for the study of spontaneous pattern
formation. Resulting from the strong coupling between cavity photons
and quantum well excitons, microcavity polaritons share the properties
of both components and, thus, display unique properties: among those,
optical and electrical injection, a high degree of tunability and
control, easy detection and direct
read-out~\cite{kavokin_laussy,Carusotto2013a,
  sanvitto_review}. Optical parametric
oscillation~\cite{baumberg00:prb}, where exciton-exciton interactions
trigger parametric scattering from a pump state to a signal state at a
lower momentum and energy and an idler state at a higher momentum and
energy, is a paradigm of polariton spontaneous pattern
formation. Here, dynamical patterns are generated by the interference
between pump, signal and idler states forming a stripe-like pattern in
real space. However, static geometrical patterns can be generated when
parametric scattering spontaneously occurs from a pump state, e.g., at
zero momentum, to two signal states at the same energy and opposite
momenta. This instability was recently realised in
triple~\cite{diederichs06} and double~\cite{Schumacher-Tignon_PRB2016}
cavities, as well as by blue-shifting the pump above the polariton
dispersion in one dimensional
cavities~\cite{abbarchi2011,ardizzone2012}. For scattering at the same
energy, scattering processes between pump and signal states of cubic
order can lead to the formation of hexagonal patterns. This was
predicted by
Refs.~\citep{Saito_PRL2013,luk2013,egorov_PRB2014,schumacher2017} and
experimentally realised in~\cite{ardizzone_2013} using a double
vertical cavity.
Alternative patterns such as vortices and vortex
lattices~\cite{Hivet_PRB2014,Boulier_SciRep2015,whittaker_2017},
vortex rings~\cite{Hamp_EPL_2015}, and solitons~\cite{Amo_Science2011}
have also been investigated in polariton quantum fluids driven by a
resonant pump. These have the additional benefit of carrying
non-trivial phase configurations and, in case of vortices, a non-zero
net angular momentum.

There is an analogy between optical patterns and Turing patterns,
where spontaneous self-organised repetitive spatial configurations
emerge out of a homogeneous distribution. Turing patterns were
first proposed in the context of chemical reactions~\cite{Turing37},
and, since then, used to describe a wide range of patterns in diverse
fields~\cite{cross93} --- such as in animal coats, skin pigmentation,
and ridges on sand dunes. The common features of Turing patterns are
non-locality, such as diffusion, and non-linear
interactions. Diffusion promotes homogeneity, yet, when the system is
driven externally by, e.g., stress, instabilities with certain
preferred wavelengths can grow exponentially because of the
non-linearities.
With a similar mechanism, Turing patterns can occur in non-linear
optical systems, such as non-linear media embedded in optical
resonators~\cite{staliunas_book}.

\begin{figure}
\centering
\includegraphics[width=1.0\linewidth,angle=0]{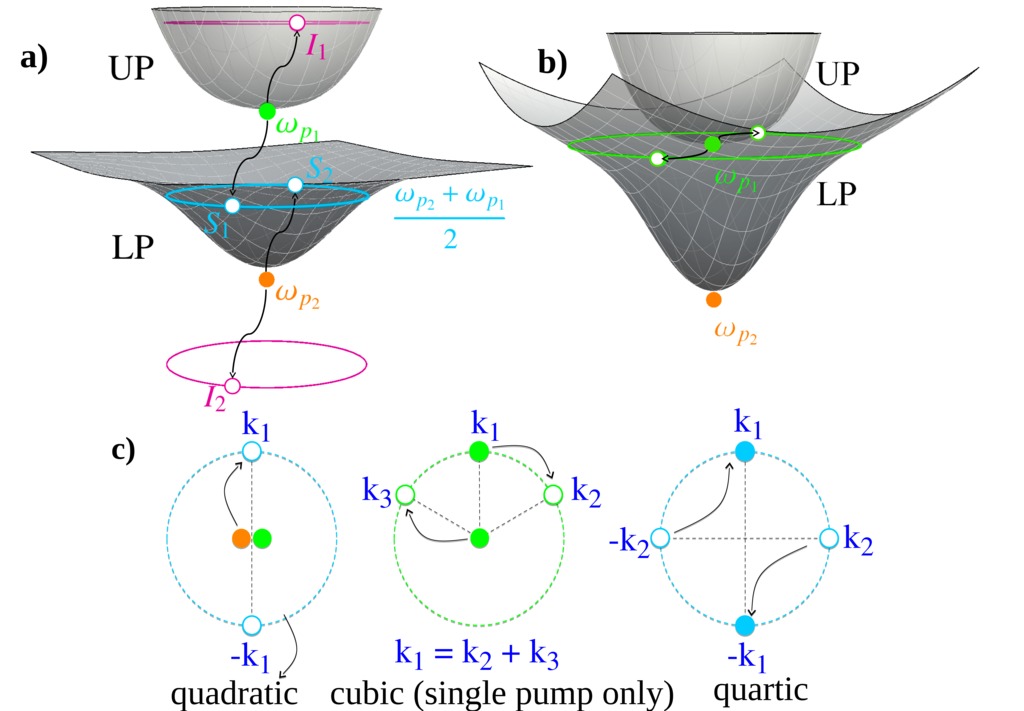}
\caption{Schematic representation of the pumping configuration and
  main scattering processes leading to two-pump (a) and single-pump
  (b) instabilities. a) Two driving laser fields are tuned at zero
  momentum almost resonantly with the upper polariton (UP) at
  $\omega_{p_1}$ (green dot) and lower polariton (LP) at
  $\omega_{p_2}$ (orange dot); UP and LP energy-momentum dispersions
  are plotted as (grey) surfaces. Interactions between the two pump
  states as well as in within each single pump state trigger
  scattering ([black] arrows) from the pump states to signal $S_{1,2}$
  (cyan) and idler $I_{1,2}$ (magenta) states. Panel b) describes the
  fixed-energy parametric scattering, which is allowed when the
  interaction renormalisation of the dispersion admits LP states at
  the pump energy $\omega_{p_1}$. c) Quadratic and quartic scattering
  processes are permitted for both two-pump (a) and single-pump (b)
  instabilities, while cubic processes are forbidden when pump and
  signal states are at different energies (a).}
\label{fig:setup}
\end{figure}

In this paper, we consider a polariton microcavity resonantly driven
by two external lasers which simultaneously pump both lower and upper
polariton branches at normal incidence so as not to explicitly break
the system translational and rotational invariance (see schematic
Fig.~\ref{fig:setup}). This pumping setup was already suggested as a
possible scheme for the generation of entangled multiple polariton
modes~\cite{Liew-Savona_PRB2011}.
More recently, a simpler but similar configuration was proposed in the
context of quantum exciton-polariton networks~\cite{liew_2018}: Here,
an inverse four-wave mixing procedure practically implements a
two-mode squeezing Hamiltonian.
However, the nature and stability of different patterns following the
spontaneous breaking of translational and rotational symmetry due to
parametric scattering has not been analysed yet.
The aim of our work is the study of those patterns that can be
generated by this pumping scheme and the control over them in terms of
the system parameters.

By complementing the results of a linear stability analysis with
numerical simulations for finite size pump profiles, we observe the
evolution between diverse patterns which can be classified as
``single-pump'' (where parametric scattering occurs at the same energy
as one of the pumps) and as ``two-pump'' (where scattering occurs at a
an energy equal to the average of the two pump energies). For two-pump
instabilities, stripe and chequerboard patterns become the dominant
steady-state solutions because cubic parametric scattering processes
are forbidden when pumps and signals are at different energies --- as
schematically depicted in Fig.~\ref{fig:setup}. This contrasts with
the single-pump case, where, because of cubic order processes,
hexagonal patterns are the most common
instabilities~\cite{Saito_PRL2013,luk2013,ardizzone_2013,egorov_PRB2014,schumacher2017}.

In a ``phase diagram'' of momentum vs. pump power, we establish the
regions of instability of the pump-only solutions, i.e., those
configurations for which only the states resonantly injected by the
external pumps are populated. At the same time, we estimate, as a
function of the pump strength, the absolute value of the momentum
typical of each instability. The values extracted from the numerical
simulations agree very well with the values found for the most
unstable modes derived within the linear stability analysis, as well
as with an estimate obtained by a simplified description of the
interaction induced renormalisation of the bare dispersion
branches. In particular, we establish that the phase diagram is
composed by different branches which can be explained in terms of both
the blue-shift and the splitting induced by the interaction between
the two-pump states mediated by excitons.
Among the two-pump instabilities, chequerboard patterns typically
occur at low pump powers. Contrary to expectations, we don't have a
clear evolution from stripes at the lowest pump powers to
chequerboards at higher pump strength. Rather, we obtain that these
instabilities alternate at low pump powers till, eventually, only
stripe solutions are allowed at very high pump powers.

If no single-pump instabilities develop, the momentum typical of these
patterns decreases monotonously as a function of the pump strength.
However, two-pump instabilities can compete with single-pump ones,
when the energy of the pump which is tuned close to the upper
polariton branch becomes resonant with the interaction renormalised
lower polariton branches. This can lead to the formation of hexagonal
patterns because of parametric scattering at the same energy, while
the system can also sustain two-pump scattering processes which
instead promote the formation of stripe and chequerboard patterns. We
can demonstrate the competition between single- and two-pump
instabilities by filtering the emission in energy, showing that the
system simultaneously undergoes different instabilities at different
energies.

Finally, we have studied the phase freedom of two-pump
instabilities. In spite of the coherent nature of the two driving
laser pumps, we demonstrate that the system is characterised by phase
freedom. In particular, the number of independent phase constraints
imposed by parametric scattering processes from the pump to the signal
and idler states is always less than the number of generated signal
and idler modes.  We show that the system spontaneously chooses the
relative phase between opposite momentum signals (which coincide with
the relative phase of opposite momentum idlers). Thus a $U(1)$ phase
symmetry is spontaneously broken in the case of stripe patterns, while
for chequerboards the phase symmetry spontaneously broken is in the
$U(1)\times U(1)$ class.

Phase freedom opens the possibility of realising macroscopic phase
coherent states and of investigating their superfluid behaviour. These
aspects have been recently analysed for optical parametric
oscillation, either by studying the current
persistence~\cite{sanvitto10,marchetti10} or by probing the system
response to the scattering against a
defect~\cite{berceanu_2015}. Further, because of the different
continuous symmetry characterising stripe and chequerboard patterns,
it would be interesting to study first order correlation functions
both in space and time so as to establish the critical behaviour of
this non-equilibrium two-dimensional system and the class of
non-equilibrium phase transition to which it
belongs~\cite{Dagvadorj_2015}.
In addition, higher order correlations would give indications of a
possible quantum
behaviour~\cite{Liew-Savona_PRB2011,Liew_NJP_2013,Kyriienko_PRB_2016,liew_2018}.

The paper is organized as follows: The model and the pumping scheme,
as well as the relevant scattering processes are introduced in
Secs.~\ref{sec:model} and~\ref{sec:twopu}. The choice of the system
parameters that are optimal for the analysis of two-pump pattern
formation is discussed in Sec.~\ref{sec:param}.  In
Sec.~\ref{sec:lires} we present the results derived within a linear
response theory, while these are compared to the results obtained with
finite size numerical simulations in Sec.~\ref{sec:numer}. We argue
about the system phase freedom in Sec.~\ref{sec:phafr}. Finally,
conclusions and perspectives form Sec.~\ref{sec:concl}.

%
\section{Model}
\label{sec:model}
We describe the dynamics of microcavity polaritons resonantly driven
by two continuous-wave laser fields shined at normal incidence
($\vect{k}_{p_1}=0 = \vect{k}_{p_2}$),
\begin{equation}
  F(\vect{r},t) = f_{p_1}(\vect{r}) e^{- i \omega_{p_1} t}
  + f_{p_2}(\vect{r}) e^{-i \omega_{p_2} t}\; ,
\label{eq:pumps}
\end{equation}
by a Gross-Pitaevskii equation for coupled cavity ($\psi_{\text{C}}^{}
(\vect{r},t)$) and exciton ($\psi_{\text{X}}^{} (\vect{r},t)$) fields
generalized to include decay and resonant pumping
($\hbar=1$)~\cite{ciuti03,carusotto04}:
\begin{equation}
  i\partial_t \begin{pmatrix} \psi_{\text{X}}^{}
    \\ \psi_{\text{C}}^{} \end{pmatrix} =
  \begin{pmatrix} 0 \\ F \end{pmatrix} +
  \left[\hat{H}_0 + \begin{pmatrix} g_{\text{X}}|\psi_{\text{X}}^{}|^2& 0 \\ 0 &
      0 \end{pmatrix}\right]
  \begin{pmatrix} \psi_{\text{X}}^{} \\ \psi_{\text{C}}^{} \end{pmatrix}\; ,
\label{eq:model}
\end{equation}
where the single polariton Hamiltonian is given by
\begin{equation}
  \hat{H}_0 = \begin{pmatrix} \omega_{\text{X},-i\nabla} - i \kappa_{\text{X}} &
    \Omega_R/2 \\ \Omega_R/2 & \omega_{\text{C},-i\nabla} - i
    \kappa_{\text{C}} \end{pmatrix} \; .
\end{equation}
Here, we assume the cavity dispersion to be quadratic,
$\omega_{\text{C},\vect{k}}=\omega_{\text{C},\vect{0}} +
k^2/(2m_{\text{C}})$, with $m_{\text{C}}=10^{-5} m_0$ ($m_0$ is the
bare electron mass),while we will neglect the exciton dispersion,
$\omega_{\text{X},\vect{k}} = \omega_{\text{X},\vect{0}}$, as the
exciton mass is much larger than the cavity photon mass, typically
$m_{\text{X}} \simeq 0.4 m_0$. Energies will be measured with respect
to the exciton one, $\omega_{\text{X},\vect{0}}$, and we define the
photon-exciton detuning as $\delta = \omega_{\text{C},\vect{0}} -
\omega_{\text{X},\vect{0}}$. Exciton and photon fields are coupled
through the Rabi splitting $\Omega_R$. The exciton-exciton interaction
is approximated as a contact interaction of strength $g_{\text{X}}$.
This approximation follows from the fact that the typical range of
exciton-exciton interaction is of the order of the Bohr radius
$a_{\text{B}} \sim$nm,
and this length is much smaller of typical polariton wavelengths $\ell
= 1/\sqrt{m_{\text{C}} \Omega_R} \sim \mu$m.
Note that the value of the interaction strength $g_{\text{X}}$ does
not influence the dynamics, as its dependence can be rescaled out from
Eqs.~\eqref{eq:model} by defining
$\tilde{\psi}_{\text{X},\text{C}}^{}= \sqrt{g_{\text{X}}}
\psi_{\text{X},\text{C}}^{}$ and $\tilde{f}_{p_{1,2}} =
\sqrt{g_{\text{X}}} f_{p_{1,2}}$. Finally, $\kappa_{\text{X}}$ and
$\kappa_{\text{C}}$ are the exciton and photon decay rates.

\subsection{Two-pump instabilities}
\label{sec:twopu}
We briefly describe in this section the main scattering processes that
characterise single- and two-pump parametric instabilities with the
scope of schematically illustrating which patterns are promoted by
each processes. The exciton-exciton interaction term inducing
scattering in the generalised Gross-Pitaevskii
equation~\eqref{eq:model} can be derived from a many-body action
written in terms of the exciton field $\psi_{\text{X}}^{}
(\vect{r},t)$:
\begin{equation}
  \mathcal{S}_{\text{int}} = \frac{g_{\text{X}}}{2} \int dt \int
  d\vect{r} \left|\psi_{\text{X}}^{} (\vect{r}, t)\right|^4 \; .
\label{eq:inter}
\end{equation}
This expression is local both in space $\vect{r}$ as well as in time
$t$, which implies that the only scattering processes allowed are
those that simultaneously conserve energy and momentum.
We assume that, aside the pump states resonantly injected by the
external lasers $(\omega_{p_1},\vect{k}_{p_1}=\vect{0})$ and
$(\omega_{p_2},\vect{k}_{p_2}=\vect{0})$, the interaction allows the
population of other states. These are indicated as signal and idler
states in the panel a) of Fig.~\ref{fig:setup}. However here, in a
simplified formulation, we assume that only signal states with energy
$\omega_{s}$ and different possible momenta $\vect{k}$ can be
populated:
\begin{equation*}
  \psi_{\text{X}}^{} (\vect{r}, t) = \sum_{i=1,2} P_{i\text{X}}^{}
  e^{-i \omega_{p_i} t} + e^{-i \omega_s t} \sum_{\vect{k}}
  S_{\vect{k}}^{} e^{-i \vect{k} \cdot \vect{r}}\; .
\end{equation*}
Assuming that only the pump states are macroscopically occupied and
perturbatively expanding in the additional signal states leads, as
explained in Sec.~\ref{sec:lires}, to the linear response theory. This
approximation scheme allows to ascertain the stability of the
pump-only solutions. However, by keeping in the expansion of the
action $\mathcal{S}_{\text{int}}$ all the terms in $S_{\vect{k}}$, we
can describe the scattering processes illustrated in panel c) of
Fig.~\ref{fig:setup} which are those responsible for the selection of
specific patterns.

In particular, the second order term
\begin{equation*}
  \mathcal{S}_{\text{int}}^{(2)} = g_{\text{X}} P_{1\text{X}}^{}
  P_{2\text{X}}^{} \sum_{\vect{k}} S_{\vect{k}}^{*} S_{-\vect{k}}^{*}
  \delta_{\omega_{p_1}+\omega_{p_2},2\omega_s} \; ,
\end{equation*}
describes the quadratic process which populates the signal energy
$\omega_s = (\omega_{p_1}+\omega_{p_2})/2$ and promotes the
population of opposite momentum states, i.e., stripe formation. This
process is allowed for both cases where either two pumps or a single
pump is present, the difference in this last case is that scattering
is only allowed when the signal states are at the same energy as the
pump.
Because of energy conservation, third order processes are allowed only
for single-pump instabilities, i.e. for
\begin{equation*}
  \mathcal{S}_{\text{int}}^{(3)} = g_{\text{X}} P_{1\text{X}}^{}
  \sum_{\vect{k}_1, \vect{k}_2, \vect{k}_3} S_{\vect{k}_1}^{}
  S_{\vect{k}_2}^{*} S_{\vect{k}_3}^{*} \delta_{\vect{k}_1,\vect{k}_2
    + \vect{k}_3} \delta_{\omega_{p_1},\omega_s} \; .
\end{equation*}
Because of the system rotational symmetry, the signal momenta
$\vect{k}_{i}$ have all to lie on the same circle, i.e., have the same
moduli. For this reason, $-\vect{k}_1$, $\vect{k}_2$ and $\vect{k}_3$
in the expression above are arranged on a equilateral triangle and,
thus, third order processes promote hexagonal patterns.
Finally, the fourth order process,
\begin{equation*}
  \mathcal{S}_{\text{int}}^{(4)} = \frac{g_{\text{X}}}{2}
  \sum_{\vect{k}_1, \vect{k}_2, \vect{k}_3, \vect{k}_4}
  S_{\vect{k}_1}^{} S_{\vect{k}_2}^{} S_{\vect{k}_3}^{*}
  S_{\vect{k}_4}^{*} \delta_{\vect{k}_1+\vect{k}_2,
    \vect{k}_3+\vect{k}_4} \; ,
\end{equation*}
populates pairs of opposite momenta states that, when arranged at
$90^\circ$, generate chequerboard, and, for any other angle, produce
rhombic patterns.
Note that this argument does not give any preference towards
chequerboard patterns with perpendicular orientation over rhombic-like
structures. However, in the following, in our numerical simulations we
will only derive squared chequerboards.

If the two pumps frequencies $\omega_{p_1}$ and $\omega_{p_2}$ are
tuned close to the upper (UP) and lower polariton (LP) branches,
respectively, then the system parameters can be chosen so as to have
the signal energy $\omega_s = (\omega_{p_1}+\omega_{p_2})/2$ relative
to two-pump processes resonant with the LP branch at a specific
momentum.
The absence of third order processes in two-pump parametric scattering
guarantees that other patterns than the hexagonal ones, such as stripe
and chequerboard, can be realised.
Interestingly, there is an analogy between the pattern formation
mechanisms described for our typically non-equilibrium system and the
theory of weak crystallisation, which is an equilibrium theory and
thus follows the principle of energy minimization.  This is briefly
discussed in the App.~\ref{app:weakc}.
In the next section, we describe the optimal choice of parameters that
leads to a large two-pump instability region.

\subsection{Choice of parameters}
\label{sec:param}
In absence of interactions, the single polariton Hamiltonian
$\hat{H}_0$ can be diagonalised in momentum space by rotating into the
lower (LP) and upper polariton (UP) basis,
\begin{equation}
  \begin{pmatrix} \psi_{\text{LP}}^{} (\k) \\ \psi_{\text{UP}}^{} (\k) \end{pmatrix}
  = \begin{pmatrix} \cos \theta_{\vect{k}} & \sin
    \theta_{\vect{k}}\\ -\sin \theta_{\vect{k}} & \cos
    \theta_{\vect{k}} \end{pmatrix} \begin{pmatrix} \psi_{\text{X}}^{}
    (\k) \\ \psi_{\text{C}}^{} (\k) \end{pmatrix} \; ,
\end{equation}
to give the LP and UP branches (see top panels of
Fig.~\ref{fig:setup}):
\begin{equation}
  \omega_{\text{LP},\text{UP},\vect{k}} =
  \Frac{\omega_{\text{X},\vect{0}} + \omega_{\text{C},\vect{k}}}{2}
  \mp \Frac{\sqrt{\left[\omega_{\text{C},\vect{k}}
        - \omega_{\text{X},\vect{0}}\right]^2 + \Omega_R^2}}{2}\; .
\end{equation}
The Rabi splitting $\Omega_R$ and the photon-exciton detuning $\delta$
determine the photon and exciton percentage that LP and UP have along
their dispersion, i.e., the Hopfield factors:
\begin{equation}
  \begin{split}
    &\cos^2 \theta_{\vect{k}}\\
    &\sin^2 \theta_{\vect{k}}
    \end{split}
    = \Frac{1}{2} \left[1 \pm \Frac{\omega_{\text{C},\vect{k}} -
        \omega_{\text{X},\vect{0}}}{\sqrt{\left[\omega_{\text{C},\vect{k}}
            - \omega_{\text{X},\vect{0}}\right]^2 +
          \Omega_R^2}}\right] \; .
\label{eq:hopfi}
\end{equation}
We rescale each overall pump strength differently, according to
\begin{align}
  f_{p_1} (\vect{r}) &= \sin \theta_{\vect{0}} F_p (\vect{r}) &
  f_{p_2} (\vect{r}) &= \cos \theta_{\vect{0}} F_p (\vect{r})\; .
\label{eq:resca}
\end{align}
This condition approximatively guarantees that, for a fixed value of
$F_p$, each pump injects the same density of UPs at $\omega_{p_1}$ and
LPs at $\omega_{p_2}$. This condition is only approximatively
guaranteed because, as soon as the pump strength is finite, UPs and
LPs are affected by different blue-shifts and thus their photon
fraction is not determined by the Hopfield factors, which instead refer
to the bare LP and UP dispersions.

We fix the Rabi splitting to $\Omega_R= 10$~meV, a value available in
GaAs-based structures. For this value, $\ell = 1/\sqrt{m_{\text{C}}
  \Omega_R} \simeq 0.58$~$\mu$m.
The other microcavity parameters, such as $\delta$ and
$\kappa_{\text{X},\text{C}}$, as well as the pump frequencies
$\omega_{p_{1,2}}$, are chosen so as to maximise the region of the
two-pump instability (see panel a) of schematic
Fig.~\ref{fig:setup}). Clearly such a scattering is not allowed at
positive detunings, for which $\omega_{\text{LP},\vect{0}} +
\omega_{\text{UP},\vect{0}} > 2 \omega_{\text{X},\vect{0}}$. We thus
choose a negative value for the detuning, $\delta=-5$~meV, which we
will see in the next section guarantees a large region of two-pump
parametric instability.

\begin{table}
\centering
\begin{tabular}{|c|c|c|}
\hline
 & case A) & case B)\\
\hline
$m_{\text{C}}$ [$m_0$] & \multicolumn{2}{c|}{$10^{-5}$}\\
\hline
$\Omega_R$ [meV] & \multicolumn{2}{c|}{10}\\
\hline
$\delta$ [meV] & \multicolumn{2}{c|}{-5}\\
\hline
$\omega_{p_1} - \omega_{\text{X},\vect{0}}$ [meV] & \multicolumn{2}{c|}{3.05}\\
\hline
$\Delta_{p_1}$ [meV] & \multicolumn{2}{c|}{-0.04}\\
\hline
$\omega_{p_2} - \omega_{\text{X},\vect{0}}$ [meV] & -9.0 & -8.5\\
\hline
$\Delta_{p_2}$ [meV] & -0.91 & -0.41\\
\hline
$\kappa_{\text{X}} = \kappa_{\text{C}}$ [meV] & 0.3 & 0.4\\
\hline
\end{tabular}
\caption{Choice of the system parameters.}
\label{tab:param}
\end{table}

The pump frequencies $\omega_{p_{1,2}}$ are chosen so as to
eliminate the possibility of bistable behaviour of each pump
separately~\cite{baas04}. When both frequencies are red-detuned just
below the UP and LP dispersion, i.e. when the pump detunings,
\begin{align}
  \Delta_{p_1} &= \omega_{p_1} - \omega_{\text{UP},\vect{0}} &
  \Delta_{p_2} &= \omega_{p_2} - \omega_{\text{LP},\vect{0}} \; ,
\label{eq:pumpd}
\end{align}
are negative, the populations of these states grow monotonically as a
function of each pump intensity, a regime known as optical limiter.

In the following, we consider two specific choices parameters
specified in Tab.~\ref{tab:param}. As explained in the next section,
by carrying on a linear response approximation, we have determined
that these parameters are optimal in order to observe two-pump
instabilities. Note that the chosen values of the exciton and photon
decay are larger than those in currently available cavities, even more
so for the state-of-the-art high-$Q$ microcavities which have been
recently grown~\cite{Snoke_PRL_2017}. Clearly, decreasing the quality
of a cavity is not difficult to achieve. As discussed in detail later
on, we find that the chosen values of the decays are optimal for
having a large region of two-pump instabilities and, at the same time,
for guaranteeing a fast convergence of two-pump patterns to a
steady-state solution.

\section{Linear response theory}
\label{sec:lires}
For homogeneous pumping, $F_p (\vect{r}) = F_p$, we can evaluate the
region of instability of the pump-only solutions by applying a linear
response approximation~\cite{whittaker05}. Here, the two-pump states,
$P_{i\alpha}$ (where $i=1,2$ indicates the pump component and $\alpha
= \text{X},\text{C}$ indicates the excitonic and photonic component)
are macroscopically occupied, while signal ($S_{i\alpha}^{}$) and
idler ($I_{i\alpha}^{}$) terms are treated perturbatively:
\begin{multline}
  \psi_{\alpha}^{} (\vect{r},t) = e^{- i \omega_{p_1} t} P_{1\alpha}^{} +
  e^{- i \omega_{p_2} t} P_{2\alpha}^{} \\ + e^{- i \omega_{p_1} t}
  \sum_{\vect{k}} \left[I_{1\alpha, \vect{k}}^{} e^{i (\vect{k} \cdot
      \vect{r} - \omega t)} + S_{1\alpha, \vect{k}}^{*} e^{- i
      (\vect{k} \cdot \vect{r} - \omega t)} \right] \\ + e^{- i
    \omega_{p_2} t} \sum_{\vect{k}} \left[S_{2\alpha, \vect{k}}^{}
    e^{i (\vect{k} \cdot \vect{r} - \omega t)} + I_{2\alpha,
      \vect{k}}^{*} e^{- i (\vect{k} \cdot \vect{r} - \omega t)}
    \right]\; .
\label{eq:fluct}
\end{multline}
The notation is the same one of panel a) of Fig.~\ref{fig:setup}: Pump
$1$ scatters at higher energy $\omega_{p_1}+\omega$ into the idler
state $I_{1\alpha}$ and at lower energy $\omega_{p_1}-\omega$ into the
signal state $S_{1\alpha}$, while pump $2$ scatters at higher energy
$\omega_{p_2}+\omega$ into the signal state $S_{2\alpha}$ and at lower
energy $\omega_{p_2}-\omega$ into the idler state $I_{2\alpha}$.

Substituting this Ansatz into the equation of
motion~\eqref{eq:model}, one obtains, at zero-th order, i.e.,
neglecting the signal and idler contributions, four mean-field
equations for the pump states~\cite{Cancellieri_mult_PRB_2011}:
\begin{equation}
  \begin{split}
  \left(\omega_{\text{X},\vect{0}}-\omega_{p_1} -i\kappa_{\text{X}} +
  G_{12} \right)P_{1\text{X}}^{} + \frac{\Omega_R}{2} P_{1\text{C}}^{}
  &=0 \\
  \left(\omega_{\text{C},\vect{0}}-\omega_{p_1}-i\kappa_{\text{C}}
  \right)P_{1\text{C}}^{}+ \frac{\Omega_R}{2} P_{1\text{X}}^{}+f_{p_1}
  &=0 \\
  \left(\omega_{\text{X},\vect{0}}-\omega_{p_2}-i\kappa_{\text{X}} +
  G_{21} \right)P_{2\text{X}}^{} + \frac{\Omega_R}{2}P_{2\text{C}}^{}
  &= 0 \\
  \left(\omega_{\text{C},\vect{0}} - \omega_{p_2}-i\kappa_{\text{C}}
  \right)P_{2\text{C}}^{} + \frac{\Omega_R}{2}P_{2\text{X}}^{}
  +f_{p_2} &=0 \; ,
  \end{split}
\label{eq:meanf}
\end{equation}
where $G_{ij}=g_{\text{X}} (|P_{i\text{X}}|^2
+2|P_{j\text{X}}|^2)$. The same equations have been already solved in
Ref.~\cite{Cancellieri_mult_PRB_2011} to demonstrate that two pumping
lasers can lead to tunable multistable hysteresis cycles with up to
three stable pump-only solutions. Here, in this work, we want to avoid
multistable regimes as they might compete with the two-pump
instability described in panel a) of Fig.~\ref{fig:setup}. For this
reason, we choose negative values for the
pump-detunings~\eqref{eq:pumpd}, a necessary condition to guarantee
the optical limiter regime. In fact, as later shown in the bottom
panels of Fig.~\ref{fig:caseD} and Fig.~\ref{fig:caseE}, for our
choice of parameters A) and B) (see Tab.~\ref{tab:param}), the pump
states $P_{i\alpha}$ grow monotonously as a function of the pump
strength $F_p$. In particular, in these figures, we plot the total
exciton density in the pump states, $n^{\text{tot}}_{\text{X}} =
|P_{1\text{X}}|^2 + |P_{2\text{X}}|^2$.

Expanding~\eqref{eq:model} to the first order in the signal and idler
terms, we can carry on a linear stability analysis of the pump only
solutions of~\eqref{eq:meanf} --- the formalism is the same as the
Erratum of Ref.~\cite{Cancellieri_errata_PRL2014}.
In this expansion, one only keeps the terms oscillating with
frequencies $\omega_{p_i} \pm \omega$, while terms oscillating with
frequencies $2\omega_{p_i}-\omega_{p_j} \pm \omega$ are neglected.
As a result, one obtains the eigenvalue equation
$\mathcal{L}_{\vect{k}} W_{\vect{k}} = \omega W_{\vect{k}}$ (diagonal
in momentum space) for the 8-component eigenvector
\begin{multline*}
  {W_{\vect{k}}}^T = \\
  (\underbrace{I_{1\text{X}, \vect{k}}, I_{1\text{C}, \vect{k}}}_{p},
  \underbrace{S_{1\text{X}, \vect{k}}, S_{1\text{C}, \vect{k}}}_{h},
  \underbrace{S_{2\text{X}, \vect{k}}, S_{2\text{C}, \vect{k}}}_{p},
  \underbrace{I_{2\text{X}, \vect{k}}, I_{2\text{C},
      \vect{k}}}_{h})\; ,
\end{multline*}
where we have explicitly indicated the particle-like ($I_{1\alpha,
  \vect{k}}$ and $S_{2\alpha, \vect{k}}$) and hole-like ($S_{1\alpha,
  \vect{k}}$ and $I_{2\alpha, \vect{k}}$) components. The Bogoliubov
matrix $\mathcal{L}_{\vect{k}}$,
\begin{equation}
  \mathcal{L} _{\vect{k}}= \begin{pmatrix} {\mathbb{L}_{\vect{k}}}_{11} &
    {\mathbb{L}_{\vect{k}}}_{12} \\ {\mathbb{L}_{\vect{k}}}_{21} &
    {\mathbb{L}_{\vect{k}}}_{22} \end{pmatrix} \; ,
\label{eq:bogol}
\end{equation}
can be decomposed in terms of the pump index block-diagonal terms
\begin{equation*}
  {\mathbb{L}_{\vect{k}}}_{ii} = \begin{pmatrix}
    E_{i\text{X},\vect{k}}^{} & \frac{\Omega_R}{2} & g_{\text{X}} P_{i
      \text{X}}^2 & 0 \\ \frac{\Omega_R}{2} &
    E_{i\text{C},\vect{k}}^{} & 0 & 0 \\ - g_{\text{X}} {P_{i
        \text{X}}^*}^2 & 0 & -E_{i\text{X},\vect{k}}^{*} &
    -\frac{\Omega_R}{2} \\ 0 & 0 & -\frac{\Omega_R}{2} &
    -E_{i\text{C},\vect{k}}^{*}\end{pmatrix} \; ,
\end{equation*}
where
\begin{align*}
  E_{i\text{X},\vect{k}}^{} &= \omega_{\text{X},\vect{k}} -
  \omega_{p_i} + 2 g_{\text{X}} n^{\text{tot}}_{\text{X}} -
  i\kappa_{\text{X}}\\ E_{i\text{C},\vect{k}}^{} &=
  \omega_{\text{C},\vect{k}} - \omega_{p_i} - i\kappa_{\text{C}}\: ,
\end{align*}
and $n^{\text{tot}}_{\text{X}}=|P_{1\text{X}}|^2+|P_{2\text{X}}|^2$,
and in terms of the off-diagonal terms:
\begin{equation*}
  {\mathbb{L}_{\vect{k}}}_{i \ne j} = 2g_{\text{X}} \begin{pmatrix}
    P_{i \text{X}} P_{j \text{X}}^* & 0 & P_{i \text{X}} P_{j
      \text{X}} & 0 \\ 0 & 0 & 0 & 0 \\ -P_{i \text{X}}^* P_{j
      \text{X}}^* & 0 & -P_{i\text{X}}^*P_{j\text{X}} & 0 \\ 0 & 0 & 0
    & 0 \end{pmatrix} \; .
\end{equation*}
The eigenvalues of $\mathcal{L}$ give the 8 branches for the complex
spectrum of excitation, $\omega_{\vect{k}}^{(n)}$. The $n=1,\dots,8$
branches can be labelled by the pump $i=1,2$ index, the excitonic and
photonic index $\alpha = \text{X},\text{C}$, and the particle-hole
$\ell=p,h$ index. In fact, in the $k\to\infty$ limit, one recovers the
rescaled exciton and photon dispersions and thus one can associate to
each of the 8 branches one specific value of these three indices
according to:
\begin{multline*}
  \lim_{k \to \infty} \omega_{\vect{k}}^{(i,\text{X},\ell)} \simeq
  \sigma_{\ell} \left[\omega_{\text{X}, \vect{k}} + 2 g_{\text{X}}
    n^{\text{tot}}_{\text{X}} -\omega_s\right. \\ \left. - \sigma_{i}\sqrt{\left(2
      g_{\text{X}} n^{\text{tot}}_{\text{X}}\right)^2 +
      \left(\frac{\omega_{p_1} - \omega_{p_2}}{2}\right)^2} \right]
  -i\kappa_{\text{X}} \; ,
\end{multline*}
where $\omega_s = (\omega_{p_1} + \omega_{p_2})/2$ and
\begin{equation*}
  \lim_{k \to \infty} \omega_{\vect{k}}^{(i,\text{C},\ell)} \simeq
  \sigma_{\ell} \left(\omega_{\text{C}, \vect{k}} -
  \omega_{p_i}\right) -i\kappa_{\text{C}} \; ,
\end{equation*}
where the sign $\sigma_{\ell}=\pm$ corresponds to the particle and
hole branches $\ell=p,h$, respectively and the sign $\sigma_i=\pm$
refers to the pump index $i=1,2$. These expressions have been derived
by neglecting the pairing terms between the particle and hole degrees
of freedom and thus are approximate and only valid at low pump
powers. Note, however, that each branch is in general characterised by
the ``particle-hole'' symmetry $\omega_{-\vect{k}}^{(n)} = -
{\omega_{\vect{k}}^{(n)}}^*$, which is a consequence of the symmetry
of $\mathcal{L}$.

\begin{figure}
\centering
\includegraphics[width=1\linewidth,angle=0]{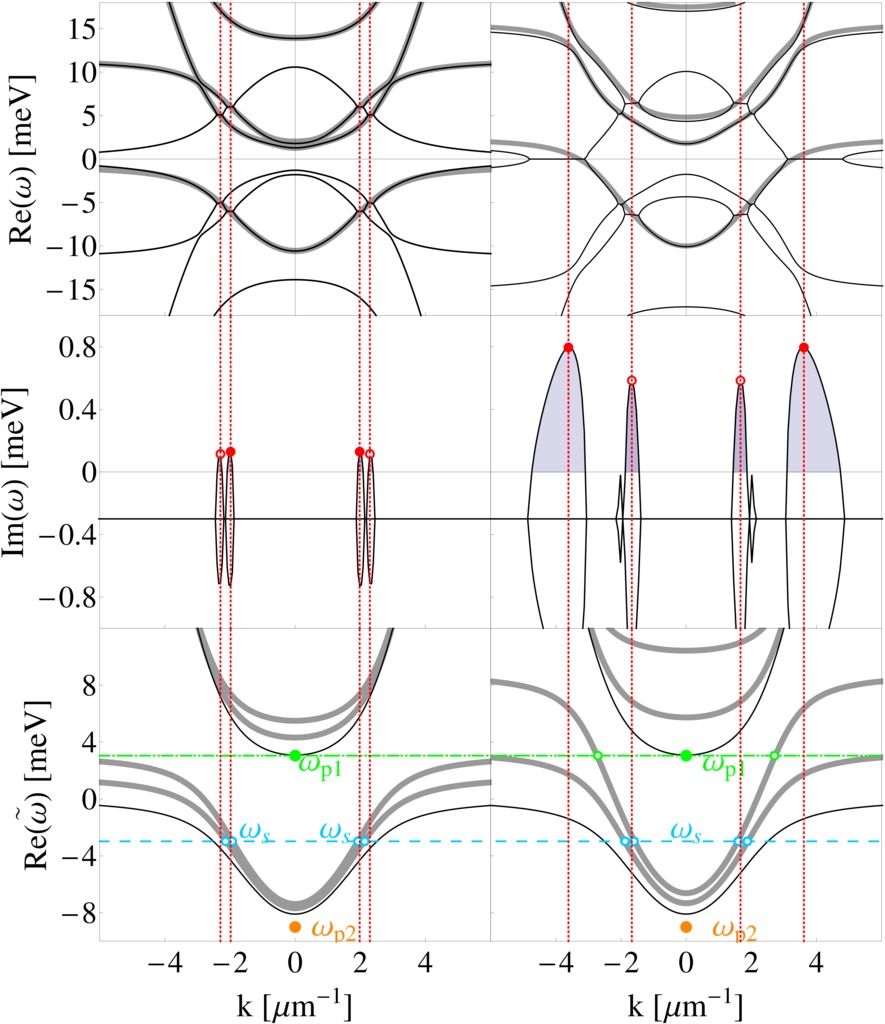}
\caption{Upper panels: The real part of the excitation spectrum
  $\Re(\omega_{\vect{k}}^{(i,\alpha,\ell)})$ (thin [black] lines),
  where its $8$ branches can be labelled by the pump $i=1,2$ index,
  the excitonic and photonic index $\alpha = \text{X},\text{C}$, and
  the particle-hole $\ell=p,h$ index (see text). The real part of the
  simplified particle-like excitation spectrum, resulting from the
  diagonalisation of~\eqref{eq:pbogo}, where we have neglected the
  particle-hole coupling terms, have been plotted as thick (grey)
  lines. Middle panels: Imaginary part of the spectrum of excitations
  $\Im(\omega_{\vect{k}}^{(i,\alpha,\ell)})$. Gray shaded regions
  indicate the unstable models for which
  $\Im(\omega_{\vect{k}}^{(i,\alpha,\ell)})>0$. The (red) vertical
  dotted lines mark the momenta of the most unstable modes. Bottom
  panels: the interaction renormalised LP and UP dispersions (thick
  [grey] lines) show that, at finite values of the pump strength, the
  interaction between the two-pump states in the particle-particle
  channel induces a blue-shift and a splitting of both the LP and the
  UP bare dispersions ([black] thin lines). The dashed (cyan) line
  indicates the energy of the signal $\omega_s=(\omega_{p_1} +
  \omega_{p_2})/2$ expected for the two-pump instability, while the
  dot-dashed (green) line indicates the pump $1$ energy $\omega_{p_1}$
  at which one can have scattering at large enough values of $F_p$
  (right panels). In all panels, the choice of parameters corresponds
  to the case A described in Tab.~\ref{tab:param}, while the pump
  strength has been fixed to $\sqrt{g_{\text{X}}}F_p=2.79$~meV$^{3/2}$
  (left panels) and to $\sqrt{g_{\text{X}}} F_p=11.18$~meV$^{3/2}$
  (right panels).}
\label{fig:spect}
\end{figure}

\begin{figure}
\centering
\includegraphics[width=1\linewidth,angle=0]{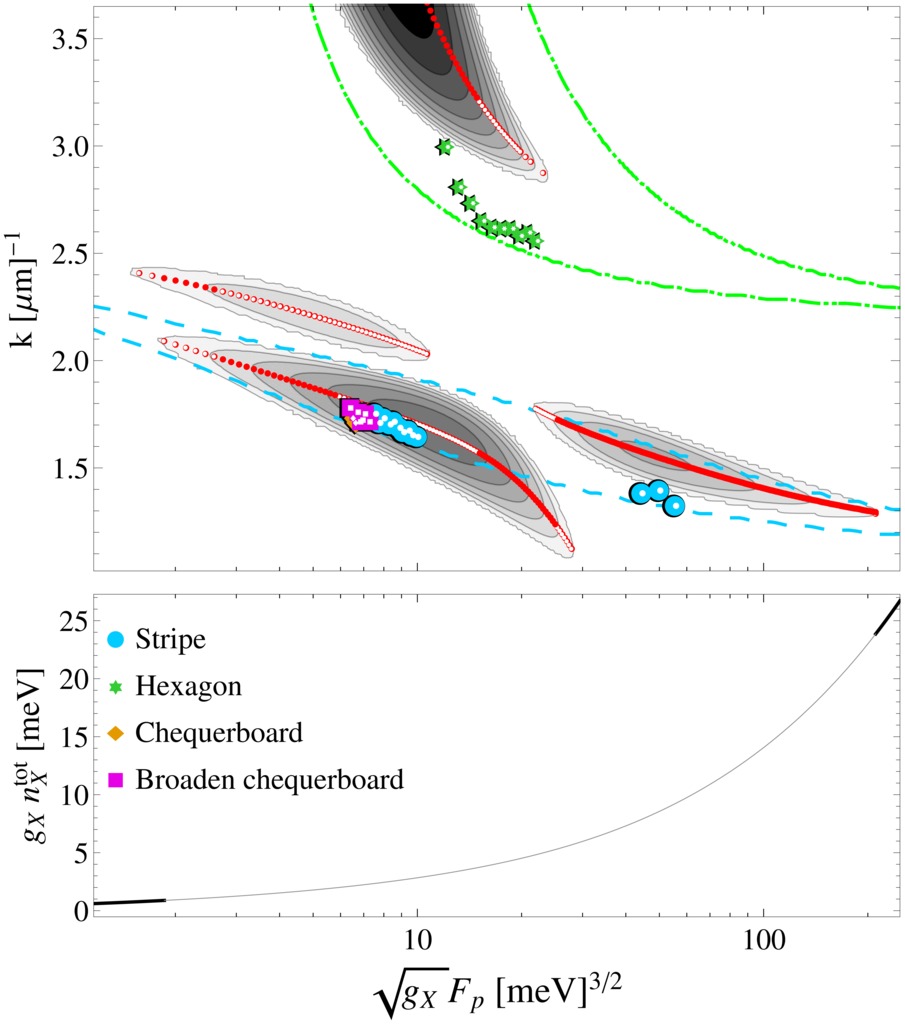}
\caption{Top panel: regions of instability of the pump-only solutions
  in the momentum $k$ and pump strength $F_p$ space.  The countourplot
  represents the region of dynamical linear instability,
  $\Im(\omega_{\vect{k}}^{(i,\alpha,\ell)})>0$. The estimates of the
  expected momenta for the two-pump instability at
  $\omega_s=(\omega_{p_1} + \omega_{p_2})/2$ ([cyan] dashed line) and
  for the single-pump instability at $\omega_{p_1}$ ([green]
  dot-dashed line) are obtained as described in the bottom panels of
  Fig.~\ref{fig:spect}. The dotted (red) lines indicate the most
  unstable modes as in the middle panels of Fig.~\ref{fig:spect} ---
  filled (red) dotted lines are those with the highest value of
  $\Im(\omega_{\vect{k}}^{(i,\alpha,\ell)})>0$. Symbols are the
  results of numerical simulations for finite size pump spots as
  described later in Sec.~\ref{sec:numer}. Bottom panel: evolution of
  the total excitonic population $n_{\text{X}}^{\text{tot}}$ of the
  pump states as a function of the rescaled pump strength. For both
  panels parameters are fixed as in the case A described in
  Tab.~\ref{tab:param}. In this case, there is an interval in pump
  strength for which two- and single-pump instabilities compete
  against each other.}
\label{fig:caseD}
\end{figure}

\begin{figure}
\centering
\includegraphics[width=1\linewidth,angle=0]{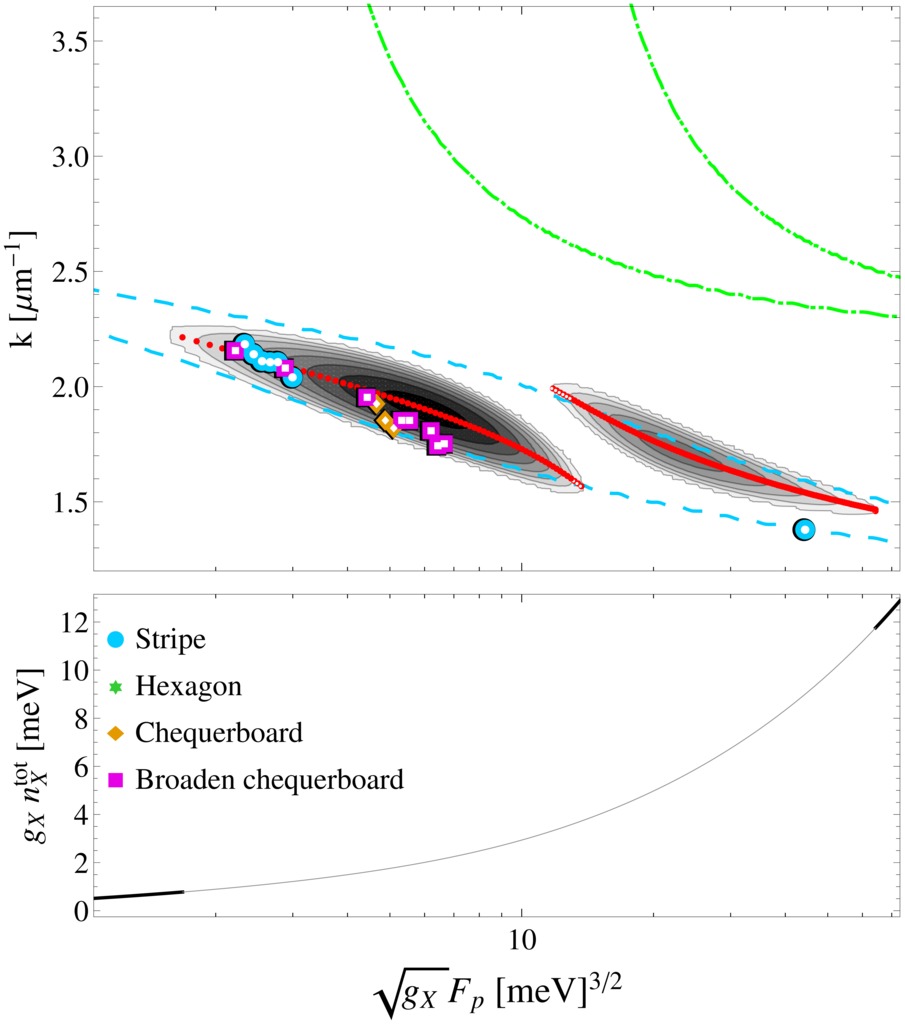}
\caption{Top and bottom panel are the same as in Fig.~\ref{fig:caseD}
  but for the system parameters fixed as in the case B described in
  Tab.~\ref{tab:param}. The parameters have been fixed so as to
  eliminate the region of instability corresponding to single-pump
  parametric scattering at the pump $1$ energy, $\omega_{p_1}$. As
  discussed later, the absence of competition between two- and
  single-pump instabilities leads to an easier convergence of stripe
  and chequerboard patterns in the numerical simulations.}
\label{fig:caseE}
\end{figure}

The real part of the excitation spectrum
$\Re(\omega_{\vect{k}}^{(i,\alpha,\ell)})$ gives information about the
renormalisation of the LP and UP bare dispersions induced by the
interaction between excitons. We plot the real part of the entire
spectrum with its 8 branches in the top panel of Fig.~\ref{fig:spect}
(thin [black] lines) for two values of the pump power --- the system
parameters for this figure correspond to the case A described in
Tab.~\ref{tab:param}.  We can observe that the ``particle-hole''
symmetry is satisfied. Further, one can observe that there are
intervals in momenta for which particle-like branches ``stick
together'' with hole-like branches. This is due to the anomalous terms
in the Bogoliubov matrix characterising the coupling between the
particle and hole degrees of freedom which induce a non-trivial (i.e.,
different from $-\kappa_{\text{X}}$ and $-\kappa_{\text{C}}$)
imaginary part of the spectrum
$\Im(\omega_{\vect{k}}^{(i,\alpha,\ell)})$ (middle panel of
Fig.~\ref{fig:spect}).

While crucial for determining the imaginary part of the spectrum, as
discussed later, these anomalous terms can be safely neglected if one
needs a simplified yet approximated information about the interaction
induced blue-shift of the LP and UP dispersions.
To show this, we consider a reduced version of the Bogoliubov matrix
where we neglect the coupling terms between the particle and hole
degrees of freedom and consider the particle-particle components only:
\begin{equation}
  \tilde{\mathbb{L}}_{\vect{k}} = \begin{pmatrix}
    E_{1\text{X},\vect{k}}^{} & \frac{\Omega_R}{2} & 2g_{\text{X}}
    P_{1 \text{X}} P_{2 \text{X}}^* & 0 \\ \frac{\Omega_R}{2} &
    E_{1\text{C},\vect{k}}^{} & 0 & 0 \\ 2g_{\text{X}} P_{2 \text{X}}
    P_{1 \text{X}}^* & 0 & E_{2\text{X},\vect{k}}^{} &
    \frac{\Omega_R}{2} \\ 0 & 0 & \frac{\Omega_R}{2} &
    E_{2\text{C},\vect{k}}^{} \end{pmatrix} \; .
\label{eq:pbogo}
\end{equation}
In the top panel of Fig.~\ref{fig:spect} we plot as thick (grey) lines
the real part of the 4 corresponding particle-like branches obtained
by diagonalising $\tilde{\mathbb{L}}_{\vect{k}}$. Note that the
imaginary part of the eigenvalues of $\tilde{\mathbb{L}}_{\vect{k}}$
consists solely of the decay terms, i.e., either $-\kappa_{\text{X}}$
or $-\kappa_{\text{C}}$, depending on the branch one refers to.
From these plots we can appreciate that, as soon as the external pumps
induce finite values of both pump fields, $P_{i\alpha}$, i.e., as soon
as $F_p\ne 0$, the interaction between the two-pump states in the
particle-particle channel, $2 g_{\text{X}} P_{1 \text{X}} P_{2
  \text{X}}^*$, induces a splitting of the LP and UP branches,
resulting in a total of four particle branches. In addition, each
branch is blue-shifted because of the $2 g_{\text{X}}
n^{\text{tot}}_{\text{X}}$ term in the diagonal components
$E_{i\alpha,\vect{k}}^{}$ of the Bogoliubov matrix. Note that,
according to the definition~\eqref{eq:fluct}, the frequency $\omega$
characterising the excitation spectrum is the frequency measured with
respect to either pump frequency $\omega_{p_{1}}$ or
$\omega_{p_{2}}$. Thus, in order to characterise the splitting and
blue-shift of the LP and UP bare dispersions (thin [black] lines in
the bottom panels of Fig.~\ref{fig:spect}) induced by the interaction,
we set to zero the terms in $\omega_{p_1}$ and $\omega_{p_2}$ in the
diagonal components $E_{i\alpha,\vect{k}}^{}$ of the simplified
Bogoliubov matrix $\tilde{\mathbb{L}}_{\vect{k}}$~\eqref{eq:pbogo} and
plot the corresponding eigenvalues $\tilde{\omega}_{\k}^{(i,\alpha)}$
in the bottom panels of Fig.~\ref{fig:spect} as thick (grey)
lines. Thus, we can quantify the splitting and blue-shift of the bare
LP and UP modes. In addition, we can describe how both splitting and
blue-shift grow when increasing the pump power $F_p$ (from the left to
the right panel).

This estimate of the interaction renormalised dispersions allows to
deduce the approximate values of the expected momenta for both
two-pump as well as single-pump instabilities. To this end, in the
bottom panels of Fig.~\ref{fig:spect} we plot as a dashed (cyan) line
the value of the two-pump signal energy $\omega_s=(\omega_{p_1} +
\omega_{p_2})/2$. Because of the splitting of the LP mode, we obtain
in this way two values of the expected two-pump instability
momenta. While at low enough pump powers (as for the left panels of
Fig.~\ref{fig:spect}), the blue-shift of the LP is not large enough to
get any LP state at the pump $1$ energy, $\omega_{p_1}$ (dot-dashed
[green] line), for big enough values of $F_p$ (right panels) the
blue-shift of at least one split LP branch is large enough to allow
single-pump parametric scattering. As discussed next, the values
obtained this way for both two-pump and single pump instability
momenta are close to those obtained evaluating the imaginary part of
the spectrum and the most unstable modes.


An additional and better estimate of the momenta at which we expect
two-pump and single-pump instabilities to occur can be obtained by
evaluating the most unstable modes. The pump-only solutions of the
mean-field equations~\eqref{eq:meanf} are stable as far as the
population of signal and idler modes in~\eqref{eq:fluct} does not grow
exponentially in time, and thus the spectrum of excitations satisfies
$\Im(\omega_{\vect{k}}^{(i,\alpha,\ell)}) < 0$. When this condition is
not met and there are values of $\k$ for which
$\Im(\omega_{\vect{k}}^{(i,\alpha,\ell)}) > 0$ for at least one of the
branches, the pump-only solutions are dynamically unstable towards the
exponential growth of these modes.
The imaginary part of the excitation spectrum is plotted in the middle
panels of Fig.~\ref{fig:spect} for two different values of the pump
strength. As observed previously, there is a trivial contribution to
the imaginary part of the spectrum coming from the decay terms: i.e.,
at large values of momenta, $\lim_{k\to
  \infty}\Im(\omega_{\vect{k}}^{(i,\alpha,\ell)})= -\kappa_{\alpha}$
(in the plot we have chosen $\kappa_{\text{X}}=
\kappa_{\text{C}}$). However, for smaller values of momentum, the
anomalous terms in the Bogoliubov matrix characterising the coupling
between the particle and hole degrees of freedom induce a non-trivial
$k$-dependent contribution to
$\Im(\omega_{\vect{k}}^{(i,\alpha,\ell)})$.
The region in momentum for which
$\Im(\omega_{\vect{k}}^{(i,\alpha,\ell)}) > 0$ characterises the
region of instability of the pump-only solutions. In
Figs.~\ref{fig:caseD} and~\ref{fig:caseE}, we plot the regions of
instabilities as a function of momentum $k$ and pump strength
$F_p$. Because of the interaction between the two-pump states and thus
the splitting of the LP and UP branches previously described, these
regions of instability are characterised by separate regions or
branches.
Note that larger decay parameters $\kappa_{\alpha}$ imply smaller
instability regions in within the linear response analysis. Thus, even
though it might seem desirable to have small values of
$\kappa_{\alpha}$ so as to obtain large regions of instability,
numerically, convergence of both two-pump and single-pump patterns
occurs more quickly for larger values of $\kappa_{\alpha}$. The
choices reported in Tab.~\ref{tab:param} as case A and B are a
compromise between these two tendencies.

The most unstable modes, i.e., those modes for which
$\Im(\omega_{\vect{k}}^{(i,\alpha,\ell)})$ is maximum as a function of
$\k$, provide information about which mode is growing faster and,
thus, about the expected momentum of the instability pattern.
In the middle panels of Fig.~\ref{fig:spect} we indicate the most
unstable modes as dotted (red) lines. The evolutions of the most
unstable mode as a function of the pump strength are plotted also as
dotted (red) lines in the top panels of both Figs.~\ref{fig:caseD}
and~\ref{fig:caseE}.
There is a good agreement between these values of instability momenta
and those previously obtained by estimating the renormalised LP and UP
dispersions. Using the same style (and color) scheme of the bottom
panels of Fig.~\ref{fig:spect}, we plot in Figs.~\ref{fig:caseD}
and~\ref{fig:caseE} as (cyan) dashed lines the expected momenta for
the two-pump instability at $\omega_s=(\omega_{p_1} + \omega_{p_2})/2$
and as (green) dot-dashed line those for the single-pump instability
at $\omega_{p_1}$.
Note that, as a consequence of the assumed isotropy of the
exciton-exciton interaction, the spectrum of excitation is also
isotropic in momentum, i.e., $\omega_{\k}^{(i,\alpha,\ell)}$ only
depends on $k=|\k|$. Thus, the most unstable modes only give
information about the ring in momentum at which the instability can
occur, but not about its direction. In other words one cannot
differentiate between stripe, chequerboard or any other pattern. In
order to obtain this information we have to carry on a full numerical
analysis, as discussed in the next section.

Finally, we note that in Fig.~\ref{fig:caseD} (parameter choice A)
there is a region in pump strength for which both two- and single-pump
instabilities are allowed and compete against each other. As discussed
in the next section, this competition between instabilities hinders
the numerical convergence of the dynamics to a steady-state. For this
reason, the parameters of case B are chosen so as to eliminate the
regions of single-pump instability, as shown in
Fig.~\ref{fig:caseE}. The absence of competition between two- and
single-pump instabilities leads to an easier convergence of patterns
in the numerical simulations, which is what we are going to discuss in
the next section.

\begin{table}
\centering
\begin{tabular}{|c|c|c|c|c|c|}
  \hline
  $\sigma_{\text{noise}}^2$ [meV$\mu$m$^2$] & 0.01 & 0.06 & 0.14 & 1.44 & 14.40 \\
  \hline
  $t_{\text{sst}}$ [ns] & 473.4 & 220.9 & 99.9 & 42.1 & 36.8 \\
  \hline
\end{tabular}
\caption{Typical times required to reach a steady-state solution
  $t_{\text{sst}}$ as a function of the white noise variance
  $\sigma_{\text{noise}}^2$ of the initial
  conditions~\eqref{eq:noise}.  These data refer to the system
  parameter choice B of Tab.~\ref{tab:param} and corresponds to the
  chequerboard pattern in Fig.~\ref{fig:squar}.}
\label{tab:noise}
\end{table}
%
\section{Numerical analysis}
\label{sec:numer}
As already mentioned in Sec.~\ref{sec:twopu}, the linear response
analysis contains only the quadratic scattering processes (see panel
c) of Fig.~\ref{fig:setup}). As such, it allows to ascertain the
stability of the pump-only solutions, and thus it provides us with the
information about the region of system parameters for which we expect
single- and two-pump instabilities to occur. However, a linear
response analysis does not permit to deduce the specific patterns
associated to each instability.
For this reason, we carry on here a full numerical analysis of the
generalised Gross-Pitaevskii equation~\eqref{eq:model} for finite size
pump spots~\eqref{eq:pumps}. In particular, in order not to break the
original rotational symmetry when both pumps are shined at normal
incidence, we consider circularly symmetric smoothed top-hat profiles
with a FWHM $\sigma_p\simeq34$~$\mu$m and a strength $F_p$ (evaluated
at the maximum value of the pump profile in real space). Note that, as
already done for the linear response theory, also in the numerical
simulations we rescale the two pump strengths according
to~\eqref{eq:resca} and thus we have a single pumping strength
parameter $F_p$ to be varied.
In order to be able to compare the numerical results with those
obtained from the linear response theory, we have chosen the same
system parameters as in the Tab.~\ref{tab:param}, and, later on, we
will report results for both parameter choices of case A) and case B).
Eq.~\eqref{eq:model} is numerically solved on a 2D grid of $N\times
N=2^8\times 2^8$ points and a separation of $0.32$~$\mu$m, in a
$L\times L = 81$~$\mu$m$\times 81$~$\mu$m box, by using a
5$^{\text{th}}$-order adaptive-step Runge-Kutta algorithm. We have
checked all our results are converged with respect to the temporal and
spacial resolution.

We impose white noise random initial conditions with zero mean,
$\langle \psi_{\alpha=\text{X},\text{C}}^{} (\vect{r},t=0)\rangle =
0$, and variance $\sigma_{\text{noise}}^2$:
\begin{equation}
  g_{\text{X}}\langle \psi_{\alpha}^{\dag}
  (\vect{r},0)\psi_{\alpha'}^{\dag} (\vect{r}',0)\rangle =
  \sigma_{\text{noise}}^2 \delta_{\alpha, \alpha'} \delta(\vect{r} -
  \vect{r}') \; .
\label{eq:noise}
\end{equation}
This is a standard procedure done in order not to bias the
steady-state solution selected by the dynamics. At the same time,
random initial conditions introduce a small explicit breaking of the
translational and rotational symmetries. This helps the numerics
evolving towards solutions for which both symmetries are spontaneously
broken, only when the system parameters are such that the symmetric
pump-only solution is unstable. We let the dynamics evolve until a
steady-state, if any, is reached and select only those solutions that
do reach a steady-state.
We have checked that none of our results depends on the choice of the
initial conditions. We find that the specific value of the noise
variance $\sigma_{\text{noise}}^2$ only affects the typical time
$t_{\text{sst}}$ the system needs to reach a steady-state. In
particular, larger values of $\sigma_{\text{noise}}^2$ typically leads
to a faster convergence in time, as shown in the Tab.~\ref{tab:noise}.
This tendency of faster convergence in time for a stronger noise in
the initial conditions is valid only for values of
$\sigma_{\text{noise}}^2$ above a certain threshold. For a weaker
noise, $\sigma_{\text{noise}}^2 < 0.01$~meV$\mu$m$^2$, we do not
observe a monotonic behaviour of $t_{\text{sst}}$.
Finally, note that the orientation of each pattern is randomly
selected. By fixing the system parameters and choosing a different
realisation of initial conditions, the system evolves exactly to the
same pattern but with a different orientation.

As previously observed in Sec.~\ref{sec:lires}, larger values of the
decay parameters $\kappa_{\alpha}$ tends to stabilise pump-only
solutions and reduce the region of instability towards spontaneous
pattern formation. However, from a numerical point of view, choosing
too small values of the decay parameters hinders the stabilisation of
a determinate pattern to a steady-state regime. A compromise between
these two behaviours has led us to the optimal values of
$\kappa_{\alpha}$ reported in Tab.~\ref{tab:param}.

Once the parameters are fixed as either in case A) or case B) of
Tab.~\ref{tab:param}, we scan through different values of the pump
strength $F_p$ and let the dynamics evolve until a steady-state is
reached for $t>t_{\text{sst}}$. As shown in Figs.~\ref{fig:caseD}
and~\ref{fig:caseE}, at either very low or very high pump powers, the
only stable solutions are those where only the two pump states at
energies $\omega_{p_1}$ and $\omega_{p_2}$ are populated, and thus no
pattern is generated. However, at intermediate pump strengths, we
observe that the pump only solutions are unstable towards the
formation of either stripe, chequerboard or hexagonal patterns. By
filtering the emission at different energies, we will later be able to
ascribe stripes and chequerboards to two-pump instabilities at an
energy $\omega_s=(\omega_{p_1} + \omega_{p_2})/2$, while hexagonal
patterns to one-pump instabilities only at an energy $\omega_{p_1}$.

\begin{figure}
\centering
\includegraphics[width=0.75\linewidth,angle=0]{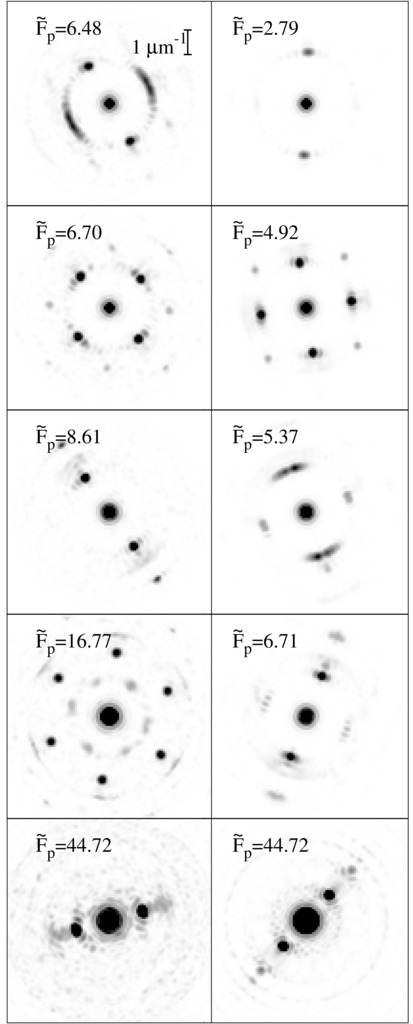}
\caption{Photon emission $|\psi_{\text{C}}^{} (\k,t)|^2$ in reciprocal
  space $\k$ ($\mu$m$^{-1}$ units) in the steady-state regime, $t >
  t_{\text{sst}}$. System parameters are those of case A of
  Tab.~\ref{tab:param} for the left column and case B for the right
  column, while the rescaled pump strength $\tilde{F}_p =
  \sqrt{g_{\text{X}}} F_p$ (meV$^{3/2}$ units) increases from the top
  panels to the bottom ones as indicated.}
\label{fig:evolu}
\end{figure}
Typical stable steady-state patterns for different values of the pump
strength are shown in Fig.~\ref{fig:evolu}, where we plot the full
photon emission $|\psi_{\text{C}}^{} (\k,t)|^2$ in momentum space $\k$
at a fixed time $t> t_{\text{sst}}$. Note that, plotting the emission
at a given time implies an integration in energy, i.e.,
$\psi_{\text{C}}^{} (\k,t) =\int d\omega e^{i\omega t}
\psi_{\text{C}}^{} (\k,\omega)$, and thus it includes the emission
from all possible energy states, including both pump states as well as
signal and idler states all emitting at different energies.
In fact, we can appreciate that all the patterns in
Fig.~\ref{fig:evolu} include emission around $\k=\vect{0}$ due to both
pump states. Emission is broaden in momentum space because of the
pumps being finite size --- note that in some panels the broadening
appears falsely increased because of the countourplot chosen
interval. In addition to the emission at zero momentum, the emission
at finite momentum characterises different patterns.
We can clearly distinguish in the patterns of Fig.~\ref{fig:evolu} a
dominant stronger emission on a momentum ring of radius
$k_{\text{primary}}$ and, in some of these panels, we can appreciate a
secondary weaker emission on a different momentum ring
$k_{\text{secondary}}$.
As analysed later, the origin of primary and secondary patterns can be
easily explained by filtering the full emission emission at different
energies.

For each pattern at a given pump strength $F_p$, we extract the value
of the primary pattern momentum $k_{\text{primary}}$ and we compare
these numerical results with the results obtained within the linear
response theory in Figs.~\ref{fig:caseD} and~\ref{fig:caseE}. Here,
the results from the numerical analysis are plotted as symbols.
We can observe that stripe patterns occur at either low pump powers
(as the panels of Fig.~\ref{fig:evolu} corresponding to $\tilde{F}_p
=2.79$~meV$^{3/2}$, $\tilde{F}_p =8.61$~meV$^{3/2}$, and $\tilde{F}_p
=6.71$~meV$^{3/2}$) or high pump powers (as for $\tilde{F}_p
=44.72$~meV$^{3/2}$). Instead, chequerboard patterns occur only at low
pump powers (as for $\tilde{F}_p =4.92$~meV$^{3/2}$ and $\tilde{F}_p
=6.70$~meV$^{3/2}$), including what we indicate as ``broaden
chequerboards'' at $\tilde{F}_p =6.48$~meV$^{3/2}$ and $\tilde{F}_p
=5.37$~meV$^{3/2}$.
It is evident in both Figs.~\ref{fig:caseD} and~\ref{fig:caseE} that
the primary pattern momentum $k_{\text{primary}}$ we extract from the
numerical simulations agrees extremely well with the lowest momentum
branch of the most unstable modes extracted from the imaginary part of
the excitation spectrum derived within the linear response theory.
If no single-pump instability is allowed, as it happens for the
parameter choice B) of Fig.~\ref{fig:caseE} and the right column of
Fig.~\ref{fig:evolu}, the primary pattern momentum
$k_{\text{primary}}$ decreases monotonously as a function of the pump
strength $F_p$. Here, there is no a clear transition from stripe to
chequerboard patterns, rather both instabilities alternate as the pump
strength increases.
However, if single-pump instabilities are allowed as for the parameter
choice A) of Fig.~\ref{fig:caseD} and the left column of
Fig.~\ref{fig:evolu}, at intermediate pump strengths, we observe the
formation of hexagonal patterns. By filtering the emission in energy
we can show that hexagonal patterns only occur at the energy
$\omega_{p_1}$ of the pump which is tuned closer to the UP. Here,
single- and two-pump instability compete against each other.

\begin{figure}
\centering
\includegraphics[width=1.0\linewidth,angle=0]{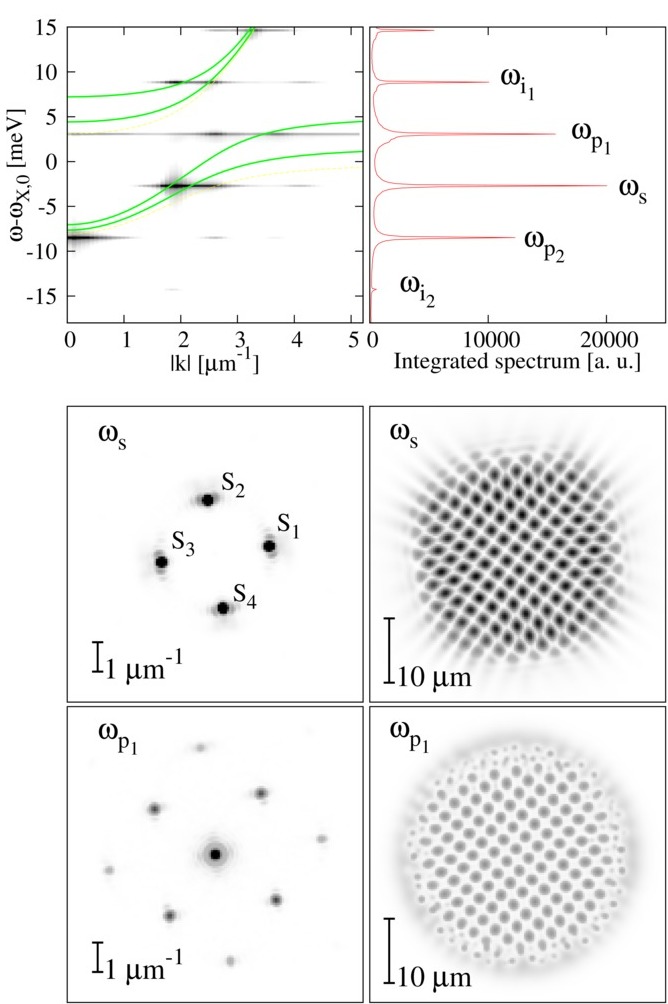}
\caption{Top left panel: Photon spectrum integrated over the momentum
  angle, $\mathcal{I}(k,\omega)=\int d\varphi |\psi_{\text{C}}^{}
  (k,\varphi,\omega)|^2$ ([grey] countourplot).  The LP and UP
  dispersions renormalised and split because of interaction effects
  are plotted as (green) solid lines: These are obtained by
  diagonalising the simplified Bogoliubov matrix~\eqref{eq:pbogo}.
  Top right panel: momentum integrated photon spectrum
  $\mathcal{I}_{\text{int}}(\omega)=\int d\k |\psi_{\text{C}}^{}
  (k,\varphi,\omega)|^2$.  The four bottom panel represent the photon
  emission momentum (left) and space (right) profiles filtered at the
  energy of the signal $\omega_s = (\omega_{p_1} + \omega_{p_2})/2$
  (top left $|\psi_{\text{C}} (\vect{k}, \omega_{s})|^2$ and top right
  $|\psi_{\text{C}} (\vect{r}, \omega_{s})|^2$) and at the energy of
  the pump $1$ $\omega_{p_1}$ (bottom left $|\psi_{\text{C}}
  (\vect{k}, \omega_{p_1})|^2$ and top right $|\psi_{\text{C}}
  (\vect{r}, \omega_{p_1})|^2$). System parameters are fixed to case
  B) of Tab.~\ref{tab:param} and the pump strength is
  $\tilde{F}_p=4.92$~meV$^{3/2}$.}
\label{fig:squar}
\end{figure}
\begin{figure}
\centering
\includegraphics[width=1.0\linewidth,angle=0]{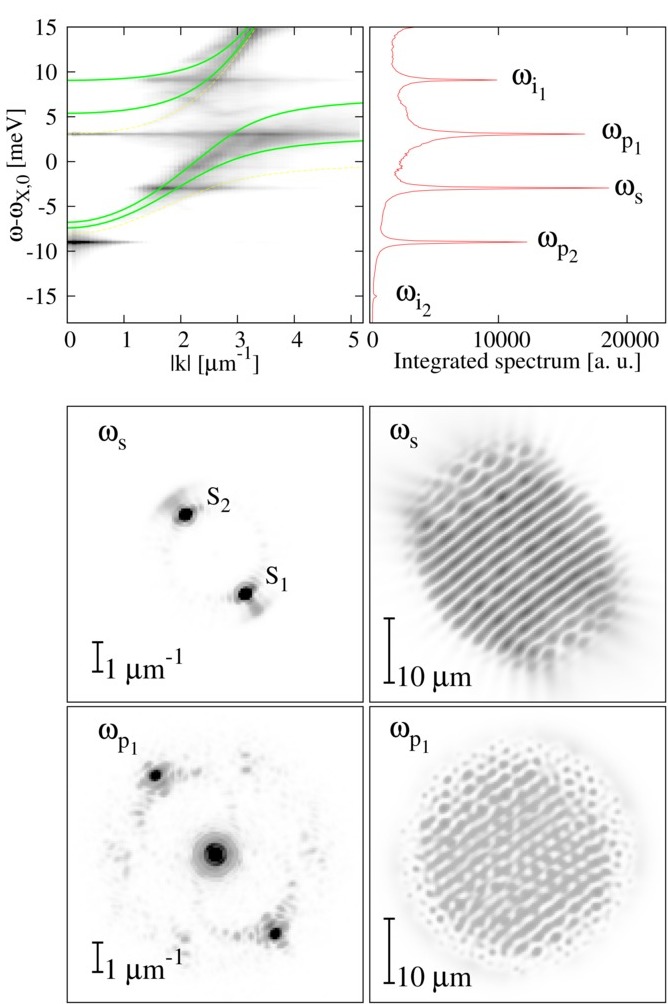}
\caption{Same as Fig.~\ref{fig:squar} for system parameters as case A)
  of Tab.~\ref{tab:param} and for a rescaled pump strength
  $\tilde{F}_p=8.61$~meV$^{3/2}$.}
\label{fig:strip}
\end{figure}
\begin{figure}
\centering
\includegraphics[width=1.0\linewidth,angle=0]{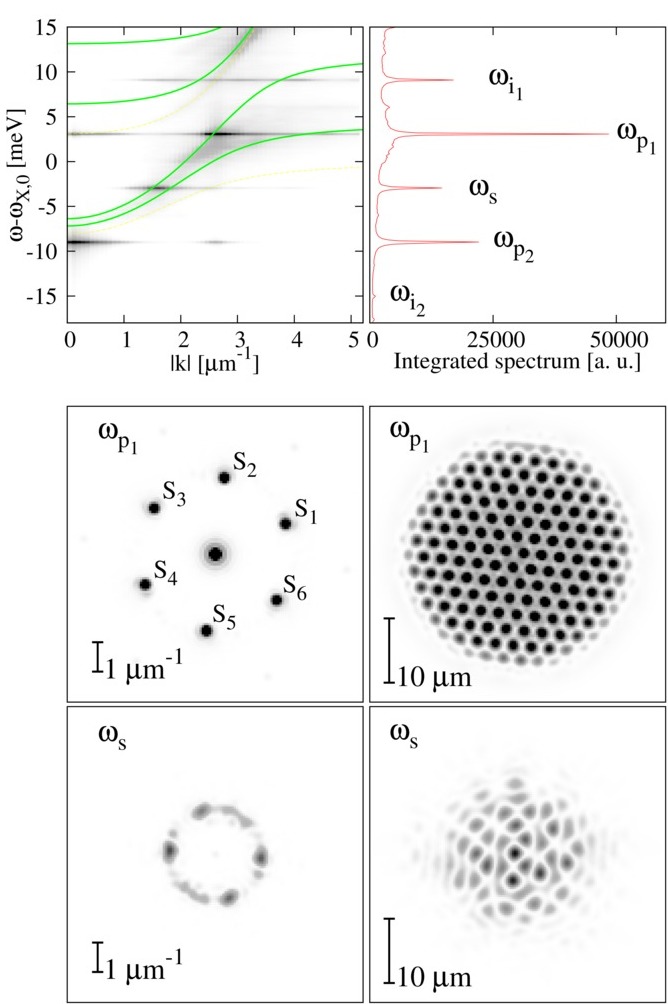}
\caption{Same as Fig.~\ref{fig:squar} for system parameters as case A)
  of Tab.~\ref{tab:param} and for a rescaled pump strength
  $\tilde{F}_p=16.77$~meV$^{3/2}$.}
\label{fig:hexag}
\end{figure}

In order to show that different patterns can appear because of
scattering processes at different energies, in
Figs.~\ref{fig:squar},~\ref{fig:strip}, and~\ref{fig:hexag} we filter
in energy the emission of typical chequerboard, stripe and hexagonal
patterns, respectively. To do this, in all these three figures we
plot, in the top left panel, the photon spectrum integrated over the
momentum angle, $\mathcal{I}(k,\omega)=\int d\varphi
|\psi_{\text{C}}^{} (\vect{k},\omega)|^2$, where $\k=(k,\varphi)$,
versus the rescaled energy $\omega-\omega_{\text{X},\vect{0}}$ and the
absolute value of momentum $k$. On the right top panel we instead plot
the momentum integrated photon spectrum
$\mathcal{I}_{\text{int}}(\omega)=\int d\k |\psi_{\text{C}}^{}
(\k,\omega)|^2$. Here, we can observe that the emission in energy is
delta-like peaked at energies equally spaced by $(\omega_{p_1} -
\omega_{p_2})/2$: aside the strong emission at the two pump energies
$\omega_{p_1}$ and $\omega_{p_2}$, we can observe the emission at the
signal energy $\omega_s = (\omega_{p_1} + \omega_{p_2})/2$ which is
the energy characteristic of two-pump instabilities. In addition, we
can appreciate a weak emission at one idler energy
$\omega_{i_1}=(3\omega_{p_1}-\omega_{p_2})/2$, above the pump $1$. The
other idler energy, $\omega_{i_2}=(3\omega_{p_2}-\omega_{p_1})/2$,
below pump $2$, is extremely weakly populated because far from being
in resonance to both the LP and the UP renormalised dispersions. This
also happens to the additional satellite states equally spaced at a
distance $(\omega_{p_1} - \omega_{p_2})/2$.

In all three figures~\ref{fig:squar},~\ref{fig:strip},
and~\ref{fig:hexag}, in the bottom four panels, we filter the emission
in energy at both the signal $\omega_s$ as well the pump $1$ energy
$\omega_{p_1}$ and plot the filtered emission both in momentum (left
panels $|\psi_{\text{C}} (\vect{k}, \omega_{s})|^2$ and
$|\psi_{\text{C}} (\vect{k}, \omega_{p_1})|^2$) as well as in space
(right panels $|\psi_{\text{C}} (\vect{r}, \omega_{s})|^2$ and
$|\psi_{\text{C}} (\vect{r}, \omega_{p_1})|^2$).
We observe that, for both cases of two-pump instabilities leading to
chequerboards (Fig.~\ref{fig:squar}) and stripes
(Fig.~\ref{fig:strip}), primary and secondary instabilities correspond
to the same pattern even if at different absolute values of momenta
and $k_{\text{primary}} < k_{\text{secondary}}$.
However, for single-pump instabilities such as the one in
Fig.~\ref{fig:hexag}, we observe that primary and secondary patterns
are not only characterised by a different absolute value of momentum
and that $k_{\text{primary}} > k_{\text{secondary}}$, in addition they
corresponds to different pattern. The stronger emission in
Fig.~\ref{fig:hexag} is at the pump $1$ energy and describe an
hexagonal pattern with $k_{\text{primary}}$. From the spectrum plotted
in the top left panel of Fig.~\ref{fig:hexag}, we can appreciate that
$k_{\text{primary}}$ is in very good agreement with the estimate we
get from the renormalised LP dispersion (green) solid line.
The emission at the signal energy $\omega_s$ is instead weaker and the
filtered emission shows a distorted chequerboard with some weaker
emission along the entire momentum ring of radius
$k_{\text{secondary}}$.

\subsection{Phase freedom}
\label{sec:phafr}
To conclude our study, we want to establish the phase freedom of
two-pump instabilities. To do this we first carry on an analytical
study valid for homogeneous pumping. Later, we compare our analytical
results with the numerical simulations for finite size pumps.

Let us start from stripe patterns. In this case, the
expansion~\eqref{eq:fluct} in signal ($S_{i\alpha}^{}$) and idler
($I_{i\alpha}^{}$) terms is limited to two opposite momentum states
$\pm \vect{k}$:
\begin{multline}
  \psi_{\alpha,\text{stripe}}^{} (\vect{r},t) = e^{- i \omega_{p_1} t}
  P_{1\alpha}^{} + e^{- i \omega_{p_2} t} P_{2\alpha}^{} \\
+ e^{- i \omega_s t} \left[S_{1\alpha}^{*} e^{i \vect{k} \cdot
    \vect{r}} + S_{2\alpha}^{} e^{-i \vect{k} \cdot \vect{r}}
  \right]\\
  + e^{- i \omega_{i_1} t} I_{1\alpha}^{} e^{-i \vect{k} \cdot
    \vect{r}} + e^{- i \omega_{i_2} t} I_{2\alpha}^{*} e^{i \vect{k}
    \cdot \vect{r}}\; ,
\label{eq:strip}
\end{multline}
where the signal energy is $\omega_s = (\omega_{p_1} +
\omega_{p_2})/2$, while the two idler energies are
$\omega_{i_1}=(3\omega_{p_1}-\omega_{p_2})/2$ and
$\omega_{i_2}=(3\omega_{p_2}-\omega_{p_1})/2$.
By substituting the expression for the stripe fields~\eqref{eq:strip}
into the equations of motion~\eqref{eq:model} and expanding to all
orders, we can infer the constraints that have to be satisfied between
the pump phases $\phi_{p_{1,2}}$ which are fixed externally and the
signal $\phi_{s_{1,2}}$ and idler $\phi_{i_{1,2}}$ phases, where
$P_{i\alpha}^{} = |P_{i\alpha}^{}| e^{i \phi_{p_{i}}}$,
$S_{i\alpha}^{} = |S_{i\alpha}^{}| e^{i \phi_{s_i}}$ and
$I_{i\alpha}^{} = |I_{i\alpha}^{}| e^{i \phi_{i_i}}$. Note that the
$\alpha=\text{X}, \text{C}$ components have their phase locked to each
other because of the $\Omega_R$ coupling in the equations of
motion~\eqref{eq:model}. We obtain that the scattering
term~\eqref{eq:inter} imposes only three independent constraints:
\begin{align*}
  \phi_{p_1} + \phi_{p_2} &= \phi_{s_1} + \phi_{s_2}\\
  2 \phi_{p_1} &= \phi_{s_1} + \phi_{i_1}\\
  2\phi_{p_2}  &= \phi_{s_2} + \phi_{i_2}\; .
\end{align*}
for the four phase terms $\phi_{s_{1,2}}$ and $\phi_{i_{1,2}}$. In
fact, the constraint for the idler phases, $\phi_{p_1} + \phi_{p_2} =
\phi_{i_1} + \phi_{i_2}$ can be obtained by the above equations and it
is not therefore independent from them.
Thus, out of the four signal and idler phases, the system is free to
spontaneously choose one relative phase only, e.g., either the
relative phase between the two signals, $\phi_{s_1} - \phi_{s_2}$, or
the one between the two idlers, $\phi_{i_1} - \phi_{i_2}$, and thus
the stripe patterns is characterised by the spontaneous breaking of a
$U(1)$ phase symmetry.

We can carry on a similar analysis for the chequerboard solution,
where we have now two pairs of opposite momenta states $\pm
\vect{k}_1$ and $\pm \vect{k}_2$, resulting into four signal states
($S_{1,2,3,4\alpha}^{}$, (see notation of Fig.~\ref{fig:squar}) and
four idler states, two of which ($I_{1,4\alpha}^{}$) at the energy
$\omega_{i_1}$ and the other two ($I_{2,3\alpha}^{}$) at the energy
$\omega_{i_2}$:
\begin{multline}
  \psi_{\alpha,\text{cheq}}^{} (\vect{r},t) = e^{- i \omega_{p_1} t}
  P_{1\alpha}^{} + e^{- i \omega_{p_2} t} P_{2\alpha}^{} + e^{- i
    \omega_s t} \\
  \left[S_{1\alpha}^{*} e^{i \vect{k}_1 \cdot \vect{r}} +
    S_{3\alpha}^{} e^{-i \vect{k}_1 \cdot \vect{r}} + S_{2\alpha}^{*}
    e^{i \vect{k}_2 \cdot \vect{r}} + S_{4\alpha}^{} e^{-i \vect{k}_2
      \cdot \vect{r}} \right]\\
  + e^{- i \omega_{i_1} t} \left[I_{1\alpha}^{} e^{-i \vect{k}_1 \cdot
      \vect{r}} + I_{4\alpha}^{*} e^{i \vect{k}_2 \cdot
      \vect{r}}\right]\\
  + e^{- i \omega_{i_2} t} \left[I_{2\alpha}^{} e^{-i \vect{k}_2 \cdot
      \vect{r}} + I_{3\alpha}^{*} e^{i \vect{k}_1 \cdot
      \vect{r}}\right]\; .
\label{eq:chequ}
\end{multline}
Substituting into the equations of motion~\eqref{eq:model} we now
obtain the following independent equations:
\begin{align*}
  \phi_{p_1} + \phi_{p_2} &= \phi_{s_1} + \phi_{s_3}\\
  \phi_{p_1} + \phi_{p_2} &= \phi_{s_2} + \phi_{s_4}\\
  2\phi_{p_1} &= \phi_{s_1} + \phi_{i_1}\\
  2\phi_{p_2} &= \phi_{s_3} + \phi_{i_3}\\
  2\phi_{p_2} &= \phi_{s_2} + \phi_{i_2}\\
  2\phi_{p_1} &= \phi_{s_4} + \phi_{i_4} \; .
\end{align*}
We get six constraints for eight phases. The system thus spontaneously
chooses two phases and is characterised by the spontaneous breaking of
a $U(1) \times U(1)$ symmetry.

The phase freedom of hexagonal patterns due to single-pump
instabilities was already derived in
Ref.~\cite{Schumacher-Tignon_PRB2016}.
Here, one has a single pump field oscillating at the energy
$\omega_{p_1}$ and with phase $\phi_p$, and six signal states, at the
same energy as the pump, which we distinguish with an index $h_j$,
with $j=1, \dots, 6$ (we assume the six signal states are arranged
clockwise, see notation of Fig.~\ref{fig:hexag}). Now the constraints
for the phases read as:
\begin{align*}
  2\phi_{p} &= \phi_{h_{j}} + \phi_{h_{j+3}}\\
  \phi_{p} + \phi_{h_j} &= \phi_{h_{j-1}} + \phi_{h_{j+1}}\; .
\end{align*}
One can easily check that out of these 12 equations only 4 are
independent, thus again giving a $U(1) \times U(1)$ phase freedom, as
in the case of the chequerboard pattern.

\begin{figure}
\centering
\includegraphics[width=1.0\linewidth,angle=0]{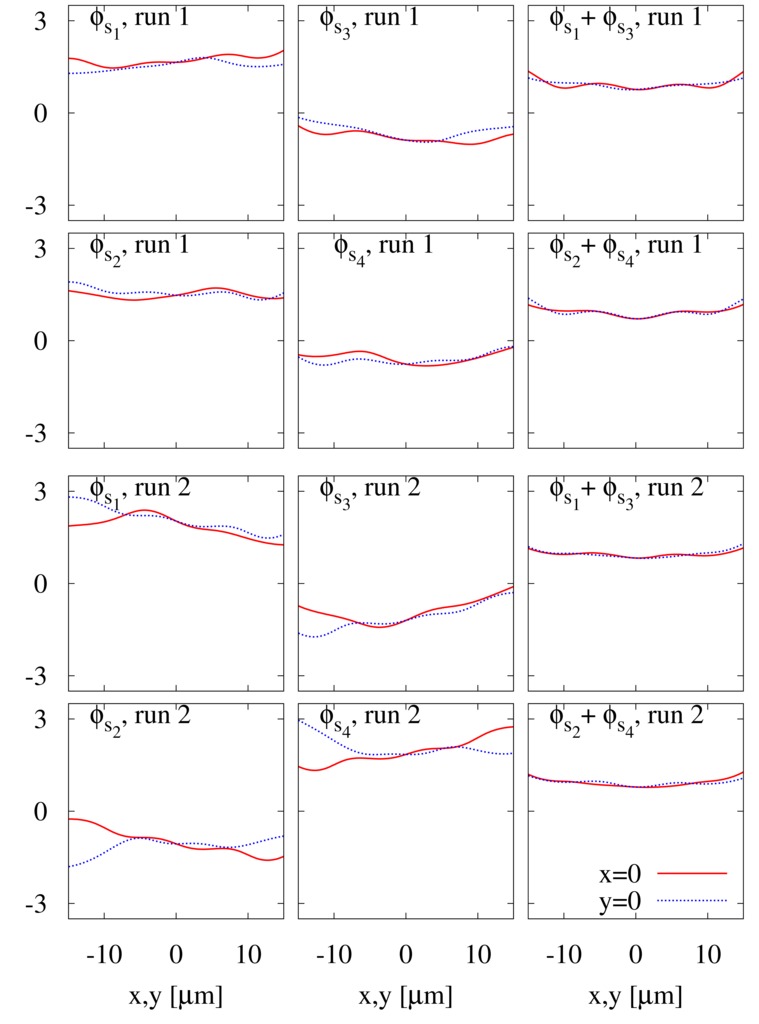}
\caption{First two columns: Extracted signal phase profiles
  $\phi_{s_{1,2,3,4}} (\vect{r})$ of the four signal states for the
  same system parameters leading to the chequerboard pattern of
  Fig.~\ref{fig:squar} and for two different noise realisations of the
  random initial conditions~\eqref{eq:noise} (labelled as run $1$ and
  run $2$). In particular, we plot the cut at $x=0$ vs $y$ ([red]
  solid line) and the cut at $y=0$ vs $x$ ([blue] dotted line).  Third
  column: We plot the sums $\phi_{s_1}(\vect{r})
  +\phi_{s_3}(\vect{r})$ and $\phi_{s_2}(\vect{r})
  +\phi_{s_4}(\vect{r})$ for the two runs in order to demonstrate the
  phase locking between them independently on the run.}
\label{fig:phases}
\end{figure}
In order to confirm the phase freedom derived analytically for
homogeneous pumping, we extract the signal phase profiles from the
finite size numerical results.
Let us refer in particular to the case of the chequerboard pattern of
Fig.~\ref{fig:squar}, yet the same procedure can be applied to any
pattern. We filter the emission at the signal energy in momentum
space, evaluating $\psi_{\text{C}} (\vect{k}, \omega_{s})$. The
amplitude of this field, for a chequerboard pattern, is peaked at four
momenta $\vect{k}_{1,2,3,4}$, all arranged on the same ring
$k_{\text{primary}}$. The phase profile $\phi_{s_{i=1,\dots, 4}}
(\vect{r})$ associated to each of these four states can be extracted
by evaluating:
\begin{multline}
  |\psi_{\text{C},i} (\vect{r}, \omega_{s})| e^{\phi_{s_{i}}
    (\vect{r})}\\
  = \sum_{\vect{k}} \psi_{\text{C}} (\vect{k}, \omega_{s})
  \theta(k_\text{cut} - |\vect{k} - \vect{k}_i|) e^{-i (\vect{k} -
    \vect{k}_i) \cdot \vect{r}} \; ,
\label{eq:phext}
\end{multline}
where we have chosen a momentum $k_{\text{cut}} \simeq
0.7$~m$\mu^{-1}$ for filtering the emission in momentum around the
four signal momenta $\vect{k}_{i=1,\dots, 4}$.
We plot the phase profiles $\phi_{s_{i=1,\dots, 4}} (x,y=0)$ and
$\phi_{s_{i=1,\dots, 4}} (x=0,y)$ of the four signal states in
Fig.~\ref{fig:phases} for the same system parameters leading to the
chequerboard pattern of Fig.~\ref{fig:squar} and for two different
noise realisations of the random initial conditions~\eqref{eq:noise}
(run $1$ and run $2$).
Even though we have subtracted in Eq.~\eqref{eq:phext} the leading
current $\vect{k}_i$ to each phase profile, we can observe in all the
panels of the first two columns of this figure that singularly all
four phases $\phi_{s_{i}} (\vect{r})$ display a residual finite
current $\vect{j}_{s_{i}} (\vect{r})= - \nabla \phi_{s_{i}} (\vect{r})
\ne \textbf{0}$ due to the system being finite size. However, we find
that these residual currents are pair-wise equal and opposite, i.e.,
$\vect{j}_{s_{1}} (\vect{r}) + \vect{j}_{s_{3}} (\vect{r}) \simeq
\textbf{0} \simeq \vect{j}_{s_{2}} (\vect{r}) + \vect{j}_{s_{4}}
(\vect{r})$, so that the phase sums $\phi_{s_{1}} (\vect{r}) +
\phi_{s_{3}} (\vect{r})$ and $\phi_{s_{2}} (\vect{r}) + \phi_{s_{4}}
(\vect{r})$ shown in the panels of the last column are almost
homogeneous in space.
Further, we observe that $\phi_{s_{1}} (\vect{r}) + \phi_{s_{3}}
(\vect{r}) \simeq \phi_{s_{2}} (\vect{r}) + \phi_{s_{4}} (\vect{r})$ within
the same run, but also between different runs, i.e. for different
random initial conditions. The fact the sums of these phases give the
same value independently on the run we consider, while the phase of
each singular phase $\phi_{s_{i}} (\vect{r})$ is different for
different runs, demonstrates the phase locking between opposite
momentum states, as well the spontaneous election of their phase
difference, i.e., the pattern phase freedom.

The specific value of the phase sums $\phi_{s_{1}} (\vect{r}) +
\phi_{s_{3}} (\vect{r}) = \phi_{s_{2}} (\vect{r}) + \phi_{s_{4}}
(\vect{r})$ would be ideally zero, as they are equal to the sum of the
two pump phases. However, as the Fourier transform from time to
frequency is evaluated numerically over a finite interval of time once
the system has evolved long enough to reach a steady-state, the
filtering process in energy produces a fictitious numerical
accumulated phase. This does not however influence our main conclusion
about phase locking and phase freedom.

Note that, as we have extracted the signal state phases, we could
similarly extract the idler phases so as to numerically check the
phase locking between $\phi_{s_1} + \phi_{i_1}$, $\phi_{s_3} +
\phi_{i_3}$, $\phi_{s_2} + \phi_{i_2}$, and $\phi_{s_4} +
\phi_{i_4}$. However, the population of the idler states, particularly
the ones at an energy $\omega_{i_2}=(3\omega_{p_2}-\omega_{p_1})/2$
below the pump $2$ energy and below the LP dispersion, is so low to
render the corresponding phases quite noisy and thus difficult to
analyse.

Finally, we have numerically extracted the phases of both stripe and
hexagonal pattern and reached a similar conclusion about phase locking
and phase freedom.

\section{Conclusions and perspectives}
\label{sec:concl}
In this paper we analyse the occurrence of Turing patterns in a
polariton microcavity which is resonantly driven by two external
lasers simultaneously pumping both lower and upper polariton
branches. The pumps are shined at normal incidence so as not to
explicitly break the system translational and rotational invariance.
We show that, by increasing the intensity of both pumps, can lead to
parametric scattering instabilities to signal states at finite
momentum, thus spontaneously breaking the system translational and
rotational symmetries. For two-pump instabilities, pumps and signals
are at different energies, and we show that stripe and chequerboard
patterns become the dominant steady-state solutions because cubic
parametric scattering processes are forbidden. This contrasts with the
case of single-pump instabilities, for which parametric scattering
occurs at the same energy as one of the two pumps. In this case, it
was already shown that hexagonal patterns are the most common
instabilities~\cite{Saito_PRL2013,luk2013,ardizzone_2013,egorov_PRB2014,schumacher2017}.
We demonstrate that our set up allows two-pump instabilities to
compete against single-pump instabilities, and that the system can
simultaneously undergo different instabilities at different energies.

Our pumping setup has been previously suggested as a possible scheme
for the generation of entangled multiple polariton
modes~\cite{Liew-Savona_PRB2011}. In that work, it was assumed that
parametric scattering would generate two signal states arranged into a
stripe configuration. Taking into account the spin-polarisation
degrees of freedom, this would generate a total of four signal states,
i.e., a square-type cluster state, for which four-mode entanglement
was demonstrated.
In our work, we analyse the nature and stability of different patterns
that can emerge from two-pump instabilities. Already without taking
into account the polarisation degrees of freedom, we show that we can
tune the system parameters so as to realise both stripe and
chequerboard patterns. This would allow the realisation of both
four-mode and eight-mode polariton entanglement if the polarisation
degrees of freedom would be taken into account.

Further, by using the pump power as a tuning parameter, we have
demonstrated that we can control the transition from stripe to
chequerboard patterns. While stripes are characterised by the
spontaneous breaking of a $U(1)$ phase symmetry, in the case of
chequerboard patterns, we show that the phase symmetry spontaneously
broken is in the $U(1)\times U(1)$ class.
We can thus tune the system across the non-equilibrium phase
transition between these two states characterised by a different
symmetry class.
This opens intriguing questions about the critical behaviour of this
non-equilibrium two-dimensional system, and the nature of the
transition from the normal phase to the ordered phase where the phase
symmetry is spontaneously broken.
It has been recently shown~\cite{Dagvadorj_2015} that polaritons
driven into the optical parametric oscillator steady-state regime
undergo a transition from a normal to a superfluid
phase~\cite{sanvitto10} that is of the Berezinskii-
Kosterlitz-Thouless type. The optical parametric oscillator regime is
characterised by the spontaneous breaking of a $U(1)$ phase
symmetry. For this parametric scattering instability, it was shown
that despite the presence of a strong drive and dissipation, the
transition from the normal to the superfluid state is governed by the
binding and unbinding of vortex-antivortex pairs, sharing similarities
to the equilibrium counterpart transition.
It is therefore natural to ask whether our stripe pattern undergoes
the same transition and to investigate the nature of the transition in
the case of chequerboard patterns. These would be the subject of
future studies.

\acknowledgments We are grateful to R. T. Brierley for contributions
at a very early stage of the project. We acknowledge useful
discussions with M. H. Szymanska and A. C. Berceanu.
Authors acknowledges financial support from the Ministerio de
Econom\'ia y Competitividad (MINECO), projects
No.~MAT2014-53119-C2-1-R and No.~MAT2017-83772-R.
%


\appendix

\section{Analogy with weak crystallisation}
\label{app:weakc}
It is interesting to note that there is some analogy between the
spontaneous appearance of a determined pattern in polariton parametric
scattering and the theory of weak
localisation~\cite{Brazovsky_JETP1987,Kats_wc_review_1993}. The
attempt to study the phase transition from a liquid to a crystal is
notoriously a hard problem to analyse which goes back to Landau, as it
implies the comparison between an infinite set of possible crystalline
and quasi-crystalline structures. However, a Ginzburg-Landau expansion
in the density modulation order parameter $\psi(\vect{r})$ can be
applied when the crystallisation transition is weakly first order,
greatly simplifying the problem.
The resulting theory of weak crystallisation assumes that the density
modulations
\begin{equation}
  \psi(\vect{r}) = \sum_{n=1}^N 2 \Re( a_n e^{i \vect{q}_n \cdot \vect{r}})
\label{eq:weakc}
\end{equation}
are small and only select a single wave-vector $|\vect{q}_n|=q_0$. The
modulated pattern, whether stripe ($N=1$), chequerboard ($N=2$),
hexagonal ($N=3$), and so on, is found by minimising a Ginzburg-Landau
type free energy functional:
\begin{equation*}
  F[\psi] = \int d\vect{r} \left\{\Frac{\tau}{2} \psi^2 + \kappa
  \left[(\nabla^2 + q_0^2)\psi\right]^2 -\Frac{\mu}{6} \psi^3 +
  \Frac{\lambda}{24} \psi^4 \right\}\; .
\end{equation*}
In two dimensions, it is easy to show that in absence of the cubic
term $\mu = 0$, there is a continuous transition from a liquid
$\psi=0$ phase for $\tau>0$ to a stripe phase ($N=1$, i.e.,
$\psi(\vect{r}) = 2 |a_1| \cos(\vect{q}_1 \cdot \vect{r} + \Phi_1)$)
for $\tau<0$ with $|a_1|=\sqrt{2 |\tau|/\lambda}$, where both phase
$\Phi_1$ and direction of $\vect{q}_1$ are randomly selected.
However, in presence of the cubic term $\mu \ne 0$, which contributes
if at least three vectors are arranged in $120^\circ$,
$\vect{q}_1+\vect{q}_2 + \vect{q}_3=\vect{0}$, the transition is to a
2D hexagonal crystal, $N=3$, with $|a_1|=|a_2|=|a_3|=|\mu| [1 +
\sqrt{1 - 10 \lambda |\tau|/\mu^2}]/(5 \lambda)$.

The free energy $F[\psi]$ is never minimised by a chequerboard $N=2$
modulation. However, for two real order parameters
\begin{flalign}
  \psi_1(\vect{r}) &= \sum_{n=1}^N 2 \Re( a_n e^{i \vect{q}_n \cdot
    \vect{r}}) &
  \psi_2(\vect{r}) &= \sum_{m=1}^M 2 \Re( b_m e^{i \vect{p}_n \cdot
    \vect{r}})\; ,
\label{eq:weak2}
\end{flalign}
with modulations in different directions $\vect{q}_n \ne \vect{p}_m$
but belonging to the same shell with momentum $q_0$, it can be shown
that the free energy
\begin{multline}
  F[\psi_1,\psi_2]= \int d\vect{r} \left\{\sum_{i=1}^{2} \kappa_i
    \left[(\nabla^2 + q_0^2)\psi_i\right]^2 \right.\\
  \left. \phantom{\sum_{i=1}^{2}} + \Frac{\tau}{2} \left(\psi_1^2 +
      \psi_2^2\right) + \Frac{\lambda}{4} \left(\psi_1^2 +
      \psi_2^2\right)^2 + \gamma \psi_1^2 \psi_2^2 \right\}
\end{multline}
undergoes a transition for $\tau<0$ from stripe $N=1$, $M=0$ when
$\gamma>\lambda/4$ to chequerboard, $N=1=M$ when $-\lambda < \gamma <
\lambda/4$. 

Note that in the formulation above we have chosen real order
parameters~\eqref{eq:weakc} and~\eqref{eq:weak2} rather than complex
ones. This choice is dictated by the phase constraints discussed in
Sec.~\ref{sec:phafr}, where we have shown that the sum of signal
phases corresponding to opposite momenta is locked to the pump phases.
Note also that in its standard formulation with local interactions,
the theory of weak crystallisation selects randomly the directions
$\vect{q}_i$ of the modulated phases. Thus, similarly to the case of
our local interaction term~\eqref{eq:inter}, the weak crystallisation
theory does not distinguish between a chequerboard arranged in a
square and the one arranged in a rhombus. However, this possibility
can be phenomenologically added by including the dependence of the
free energy on the angles between ordering wave vectors.


%

\end{document}